\documentclass[12pt]{iopart}
\usepackage{graphicx}
\usepackage{hyperref}
\usepackage{cite}
\begin{document}

\title[Geometric Resonances in BECs with Two- and Three-Body Interactions]{Geometric Resonances in Bose-Einstein Condensates with Two- and Three-Body Interactions}

\author{Hamid Al-Jibbouri$^1$, Ivana Vidanovi\' c$^{2,3}$, Antun Bala\v z$^{2,4}$ and Axel Pelster$^{4,5}$}

\address{$^1$Institut f\"ur Theoretische Physik, Freie Universit\"at Berlin, Arnimallee 14, 14195 Berlin, Germany\\
$^2$Scientific Computing Laboratory, Institute of Physics Belgrade, University of Belgrade, Pregrevica 118, 11080 Belgrade, Serbia\\
$^3$Institut f\"ur Theoretische Physik, Johann Wolfgang Goethe-Universit\"at, 60438 Frankfurt am Main, Germany\\ 
$^4$Hanse-Wissenschaftskolleg, Lehmkuhlenbusch 4, 27733 Delmenhorst, Germany\\
$^5$Fachbereich Physik und Forschungszentrum OPTIMAS, Technische Universit\"at Kaiserslautern, 67663 Kaiserslautern, Germany}
\ead{antun@ipb.ac.rs}
\begin{abstract}
We investigate geometric resonances in Bose-Einstein condensates by solving the
underlying
time-dependent Gross-Pitaevskii (GP) equation for systems with two-and
three-body interactions in an axially-symmetric harmonic trap. To this end, we
use a recently developed analytical method [Vidanovi\' c I \etal 2011 \PR A {\bf 84} 013618],
based on both a perturbative expansion and a Poincar\'e-Lindstedt
analysis of a Gaussian variational approach, as well as a detailed numerical
study of a set of ordinary differential equations for
variational parameters. By changing the anisotropy of the
confining potential, we numerically observe and analytically describe strong
nonlinear effects: shifts in the frequencies and mode coupling of collective modes,
as well as resonances. Furthermore, we discuss in detail
the
stability of a Bose-Einstein condensate in the presence of an attractive
two-body interaction and a repulsive three-body interaction.
In particular, we show that a small repulsive three-body interaction is able to significantly extend the
stability region of the condensate.
\end{abstract}
\pacs{03.75.Kk, 03.75.Nt, 67.85.De}

\section{Introduction}

The experimental discovery of Bose-Einstein condensation \cite{bec1,bec2,bec3,bec4,bec5,bec6} has instigated extensive experimental and theoretical studies of
ultracold atoms and molecules. In particular, many experiments have
focused on collective excitations of harmonically trapped
Bose-Einstein condensates (BECs), as
their frequencies can be measured to the order of a few per mill \cite{3a,3b,Stamper-Kurn,3c}
and calculated analytically \cite{5,5a,05,5b,5c,5e,8},
and thus provide a reliable method for
extracting ultracold system parameters.

A wide variety of interesting nonlinear phenomena are observed in
collective excitations of BECs, including frequency shifts \cite{c9,9g}, mode
coupling \cite{c9,c10,c11}, damping
\cite{Stamper-Kurn,c12}, nonlinear interferometry \cite{Chaohong}, as well as collapse and
revival of oscillations \cite{c9,c14,c15}.
The collective oscillation modes can be induced in a BEC by modulating the
external potential trap
\cite{3a,3b,9b,9c,c9,9d,9hh1,9hh2,99d,9hh3,9hh4,99hh1,99hh4,99hh2,99hh3,99hh5,99hh6},
the s-wave scattering length \cite{9g,gaul1,9f,9h,gaul2} or three-body interactions \cite{9h,9gg}.

Resonant coupling between
collective modes in a BEC was experimentally observed \cite{c10,c10a}, and it was shown
that, when the parity quadrupole mode is excited by changing the trap anisotropy
parameter above a certain value, it is possible to achieve an energy transfer
between
modes at a rate~\cite{c11} which is comparable to the collective mode frequency. In
reference~\cite{c9}, the frequency shift of collective modes due to the trap anisotropy in a generic axially-symmetric geometry
was studied, and it was shown that the
collective modes exhibit a resonant behaviour for specific
values of the trap anisotropy, which are called geometric resonances, and that the strong
effects can be observed even for oscillations of relatively small amplitude.
The excitations and coupling of quadrupole and scissors modes
in two-component BECs were investigated in reference~\cite{3bb}. Recently,
also a coupling of the dipole, breathing and quadrupole mode close to a Feshbach resonance was analysed in reference~\cite{Bagnato12}.

In this paper we study geometric resonances and resonant mode coupling in BECs with two- and three-body contact interactions.
Theoretical studies of collective excitations are usually focused on two-body contact interactions
due to the diluteness of quantum gases \cite{3c,9f,9g,pi,c9,c11,c10}.
However, the experimental progress with BECs in atomic waveguides and on the surface of
atomic chips, which involve a strong increase in the density of BECs, necessitates also the study of three-body interactions \cite{10g,10h,10j}.
Theoretical and experimental
studies
\cite{10g,10o,10p} for a BEC of ${}^{87}$Rb atoms indicated that the real
part of the three-body interaction term can be
$10^3-10^4$ times larger than the imaginary part. The imaginary part, which
arises from three-body recombinations, limits the lifetime of the
condensate. However, even for a small strength of the three-body interaction, the region
of stability for the condensate can be extended considerably according to references~\cite{11n,10q,10r,11f}.
We study this in more detail and provide a phase diagram which demonstrates the significantly enhanced stability of BECs due to three-body interactions.

Due
to the three-body interaction, the density profile \cite{10r}, the excitation spectrum of the collective oscillations \cite{10t,10v},
as well as the modulation instability  of a trapped BEC \cite{11a} is modified. The
effects of the three-body interaction were furthermore studied in ultracold bosonic atoms in an optical lattice
\cite{10j,optical1,optical2,opteical3,opteical4,opteical5,sowinski,FatkhullaKh.Abdullaev,Chen},
BCS-BEC crossover \cite{Dasgupta}, complex solitons BEC \cite{Roy} and vortex
BEC \cite{Juan}. In addition, an extensive work was done on the study of cubic-quintic non-linear equations, most notably in the context of nonlinear optics and superfluid helium.
Even though these studies were done in uniform systems, many of the results are quite relevant for trapped systems as well.
In particular, we mention studies of cavitation \cite{Josserand1}, droplets \cite{Josserand2}, as well as dynamics, solitary waves, and vortex nucleation \cite{Berloff}.
The transition temperature, the depletion of the
condensate atoms, and the collective excitations of a BEC with two- and
three-body interactions in an
anharmonic trap at finite temperature are studied in reference~\cite{11}. Reference~\cite{11c} shows that the frequency of the collective excitation is also
significantly affected by the strength of the
three-body interaction and the anharmonicity of the potential. In
reference~\cite{11d} the
authors investigated the collective excitations and the stability of a BEC in a
one-dimensional trapping geometry for the case of repulsive or attractive three-body and repulsive
two-body interactions.

Motivated by this, we study here the dynamics of the condensate with both two- and three-body contact interactions
in general and its collective oscillation modes in particular by changing the geometry
of the trapping potential.
Within a Gaussian variational approach, the
partial differential equation of Gross and Pitaevskii is transformed in section~\ref{sec:VA} into a set
of ordinary differential
equations for the condensate widths.
We then
discuss in detail in section~\ref{sec:SD} the resulting stability of the
condensate. First, we
consider the case of an attractive two-body interaction and a
vanishing
three-body interaction, and afterwards the case of
attractive two-body and repulsive three-body
interaction. In section~\ref{sec:shifts} we study geometric resonances
and
derive an explicit analytic results for the frequency shifts for the case of an
axially-symmetric condensate based on
a perturbative expansion and a Poincar\'e-Lindstedt method. This frequency shift is
calculated for a quadrupole mode in subsection~\ref{subsec:QM}, for a breathing mode
in subsection~\ref{subsec:BM}, and the derived analytical results are then compared
with the results of numerical simulations in subsection~\ref{subsec:numerics}. In that
subsection we also compare results of numerical simulations for radial and
longitudinal condensate widths and the corresponding excitations spectra
with the analytical results obtained using perturbation theory. Then, in
section~\ref{sec:MC}, we analyse the
resonant mode coupling and the generation of second harmonics of the
collective modes. Finally, in section~\ref{sec:Con} we
summarise our
findings and present our conclusions.

\section{Variational Approach}
\label{sec:VA}

The dynamics of a Bose-Einstein-condensed gas in a trap at zero temperature is
well
described by the time-dependent GP equation \cite{gross,pitaevskii,pitstri-book,petsmi-book,11c,11d}. Usually, only
two-body
contact interactions are considered due to the diluteness of the gas. In this
paper, however, we study systems where also three-body
contact interactions have to be taken into account \cite{dalfovormp,10o}. In
that case, the GP equation has the form
\begin{equation}
\hspace*{-10mm}
 i\hbar\frac{\partial}{\partial t}\psi({\bf r},t) = \Big[-\frac{\hbar^{2}}{2
m}\Delta+V({\bf r})+ g_2 N \left|\psi({\bf r},t)\right|^{2} 
+ g_3 N^2\left|\psi({\bf r},t)\right|^{4}\Big] \psi({\bf r},t)\, .
\label{eq:gp}
\end{equation}
where $\psi({\bf r},t)$ denotes a condensate wave function normalised to unity,
and $N$ is the total number of atoms in the condensate. On the right-hand side
of the above equation
we have a kinetic energy term, an external axially-symmetric harmonic trap
potential
$V({\bf r})=\frac{1}{2} m \omega_{\rho}^2 \left(\rho^2+ \lambda^2 z^2\right)$
with the anisotropy parameter $\lambda=\omega_z/\omega_\rho$, while the
parameters $g_2$ and $g_3$
account for the strength of two-body and three-body contact interactions,
respectively. The two-body interaction strength $g_2=4 \pi \hbar^2  a/m$
is proportional to the $s$-wave scattering length $a$, where $m$ denotes the mass
of the corresponding atomic species.

The three-body interaction strength $g_3$ becomes important for large values of the
$s$-wave scattering length,
but also for small values of $a$ close to the ideal gas regime. It is well
known that
the stability against the collapse of $^{85}$Rb cannot be described by using only the
two-body scattering \cite{altin}. The three-body scattering also plays an essential role
in understanding the Efimov physics, where three bosons form a bound state \cite{efimov1,efimov2}. Braaten and Nieto
\cite{braatennieto} have used an
effective field theory to calculate the strength of the three-body interaction, which effectively arises from the two-body interaction, and obtained the result
$g_3(\kappa)=384\pi(4\pi-3\sqrt{3})[\ln \kappa a+B]\hbar^2a^4/m$, where $\kappa$
is an arbitrary
wave number and $B$ is a complex constant, both being suitably fixed in
reference~\cite{braatennieto}. Thus, in general, the effective three-body coupling strength represents a complex number,
where its
imaginary part describes recombination effects. However, its real
part turns out to be much larger, and the fit to experimental data for $^{85}$Rb and
$^{87}$Rb gives typical
values for Re$(g_3)/\hbar$ of the order of $10^{-27} {\rm {cm^{6} s^{-1}}}$ to
$10^{-26} {\rm {cm^{6} s^{-1}}}$ \cite{11e,11,bulgac}.

Equation (\ref{eq:gp}) can be cast into a variational problem, which corresponds
to the extremization of the action defined by the Lagrangian
$L(t)=\int \mathcal{L}({\bf r},t)\, d {\bf r}\, ,$
with the Lagranian density
\begin{equation}
\hspace*{-15mm}
\mathcal{L}({\bf r},t)=\frac{i \hbar}{2}\left(\psi^* \frac{\partial
\psi}{\partial
t}- \psi\frac{\partial \psi^*}{\partial t} \right)- \frac{\hbar^2}{2m}|\nabla
\psi|^2-V({\bf r}) |\psi|^2  -
\frac{g_2 N}{2}
|\psi|^4
-\frac{g_3 N^2}{3} |\psi|^6\, .
\label{eq:lag}
\end{equation}
In order to analytically study the dynamics of BEC systems with two- and
three-body interactions, we
use the Gaussian variational ansatz, which was introduced in references~\cite{5c,8}.
For an axially symmetric trap, this time-dependent
ansatz reads
\begin{equation}
\hspace*{-10mm}
\psi^{\rm G} (\rho,z,t)={\mathcal N}(t)
\exp\left[-\frac{1}{2}\frac{\rho^2}{u_{\rho}(t)^2}+i
\rho^2 \phi_{\rho}(t)\right] 
\exp\left[-\frac{1}{2}\frac{z^2}{u_z(t)^2}+i z^2
\phi_ z (t)\right],
\label{eq:G}
\end{equation}
where $\mathcal{N}(t)=1/\sqrt{\pi^{\frac{3}{2}}
u_{\rho}^2(t) u_z(t)}$ is a normalisation factor, while $u_{\rho}(t)$, $u_{z}(t)$,
$\phi_z(t)$, and $\phi_{\rho}(t)$ are variational parameters, representing
radial and axial condensate widths and the corresponding phases.
The ansatz (\ref{eq:G}) describes dynamics of the condensate in terms of the
time-dependent condensate widths and phases, while no center-of-mass
motion is considered here. A similar variational ansatz including the center-of-mass motion
has been studied in reference \cite{trombettoni}, and would be suitable
to investigate how the center-of-mass motion couples to the collective oscillation modes in the presence of three-body interactions.

If we insert the Gaussian ansatz (\ref{eq:G}) into the Lagrangian~(\ref{eq:lag}), we
obtain the Lagrange function
\begin{eqnarray}
\label{eq:lagf}
\hspace*{-18mm}
L(t)&=&-\frac{\hbar}{2}\left(2\dot \phi_{\rho} u_{\rho}^2 +\dot \phi_z
u_z^2\right)-\frac{m \omega_\rho^2}{2}\left(u_{\rho}^2+\lambda^2
\frac{u_z^2}{2}\right) \\
\hspace*{-18mm}
&& -\frac{\hbar^2}{2
m}\left[\left(\frac{1}{u_{\rho}^4}+4\phi_{\rho}^2\right)u_{\rho}^2+\left(\frac{1
}{u_z^4}+4\phi_z^2\right)\frac{u_z^2}{2}\right]  -
\frac{g_2 N}{2(2 \pi)^{3/2} u_{\rho}^2 u_z}
-\frac{g_3 N^2}{9 \sqrt{3} \pi^3 u_{\rho}^4 u_z^2}\, .\nonumber
\end{eqnarray}
From the corresponding Euler-Lagrange equations we obtain the equations of
motion for all
variational parameters.
The phases $\phi_{\rho}$ and $\phi_z$ can be expressed explicitly in
terms of first derivatives of the widths $u_{\rho}$  and $u_z$ according to
\begin{equation}
\phi_{\rho} = \frac{m \dot u_{\rho}}{2 \hbar u_{\rho}}\, ,
\qquad
\phi_z  = \frac{m \dot u_z}{ 2 \hbar u_z}\, .
\label{eq:phizu}
\end{equation}
Inserting equations~(\ref{eq:phizu}) into the Euler-Lagrange equations for the widths, we get
second-order differential equation for $u_\rho$ and $u_q$. After introducing the
dimensionless parameters
\begin{equation}
\tilde\omega_i = \omega_i/\omega_{\rho}\, ,\qquad
\tilde u_i = u_i /\ell\, , \qquad
\tilde t =\omega_{\rho}\, t\,
\end{equation}
with the oscillator length  $\ell=\sqrt{\hbar/m\omega_\rho}$,
we obtain a system of two second-order differential equations
for $u_{\rho}$ and $u_z$ in the dimensionless
form
\begin{eqnarray}
\hspace{-0.5cm}
\ddot{u}_{\rho}+u_{\rho}-\frac{1}{u_{\rho}^3}-\frac{p}{u_{\rho}^3
u_z}- \frac{k}{u_{\rho}^5
u_z^2}&=&0\, ,\label{eq:ur}\\
\hspace{-0.5cm} \ddot{u}_z+\lambda^2
u_z-\frac{1}{u_z^3}-\frac{p}{u_{\rho}^2
u_z^2} - \frac{k}{u_{\rho}^4
u_z^3}&=&0\, ,\label{eq:uz}
\end{eqnarray}
where, for simplicity, we drop the tilde sign in the dimensionless widths. In
the above equations,
\begin{equation}
p= \frac{g_2 N}{(2 \pi)^{3/2} \hbar \omega_\rho \ell^3}=\sqrt{\frac{2}{\pi}}\, \frac{Na}{\ell}
\end{equation}
denotes the dimensionless two-body interaction strength, while the parameter
\begin{equation}
k=\frac{4 g_3 N^2}{9 \sqrt{3} \pi^3 \hbar\omega_\rho\ell^6}
\end{equation}
is the dimensionless three-body interaction strength,
which can be also expressed in terms of $p$ as
\begin{equation}
k=\frac{32}{9 \sqrt{3}}\, \frac{g_3 \hbar \omega_\rho}{g_2^2} p^2\, .
\label{kp}
\end{equation}
For $N=10^5$ atoms of $^{87}{\rm Rb}$ \cite{c10a,10j} in a trap with
$\omega_{\rho}=2 \pi\times112 \ {\rm Hz}$, the two-body interaction
strength is $g_2= 5\hbar \times 10^{-11} {\rm {cm^{3} s^{-1}}}$, yielding
$p=426$. The three-body interaction is of the order $g_3\approx \hbar \times
10^{-26}
{\rm {cm^{6} s^{-1}}}$ \cite{10j}, which gives the dimensionless three-body
interaction value $k=1050$.

Although the value of $k$ is larger than that of $p$, the corresponding terms in
equations~(\ref{eq:ur}) and (\ref{eq:uz}), i.e.
$k/u_{\rho}^5 u_z^2$ and $k/u_{\rho}^4 u_z^3$,
are suppressed by the factor $u_\rho^2u_z$ compared to the respective $p$-terms.
The value of this factor can be estimated by taking into account the equilibrium
positions $u_{\rho 0}$ and $u_{z0}$, which are obtained by solving the
stationary equations
\begin{eqnarray}
u_{\rho 0}&=&\frac{1}{u_{\rho 0}^3}+\frac{p}{u_{\rho 0}^3 u_{z0}}+
\frac{k}{u_{\rho 0}^5 u_{z0}^2}\, , \label{eq:ur0}\\
\lambda^2 u_{z0}&=&\frac{1}{u_{z0}^3}+\frac{p}{u_{\rho 0}^2 u_{z0}^2}
+\frac{k}{u_{\rho 0}^4 u_{z0}^3} \label{eq:uz0}\, .
\end{eqnarray}
For the anisotropy $\lambda=3/2$, one numerically obtains $u_{\rho 0}\approx
3.69$ and $u_{z0}\approx 2.47$, yielding the value $u_{\rho 0}^2u_{z0}\approx
33.6$. This shows
that the terms proportional to $k$ have the effective coupling $k/33.6\approx
31.2$, which makes them small corrections of the order of 7\% to the leading
two-body interaction
terms. However, if the system exhibits resonances, this may no longer be true
anymore, and three-body interactions can play a significant role for the system
dynamics. In this paper
we study geometric resonances,
where it turns out to be necessary to take into account effects of three-body
interactions. The $s$-wave scattering length can be tuned to any value, large or small,
positive or negative, by applying an external magnetic field, using the Feshbach
resonance technique \cite{Grimm,Song}.
Therefore, in this paper we will consider a range of experimentally realistic
values for dimensionless interaction strengths $p$ and $k$.

\begin{figure}[!t]
\centering
\includegraphics[height=5cm]{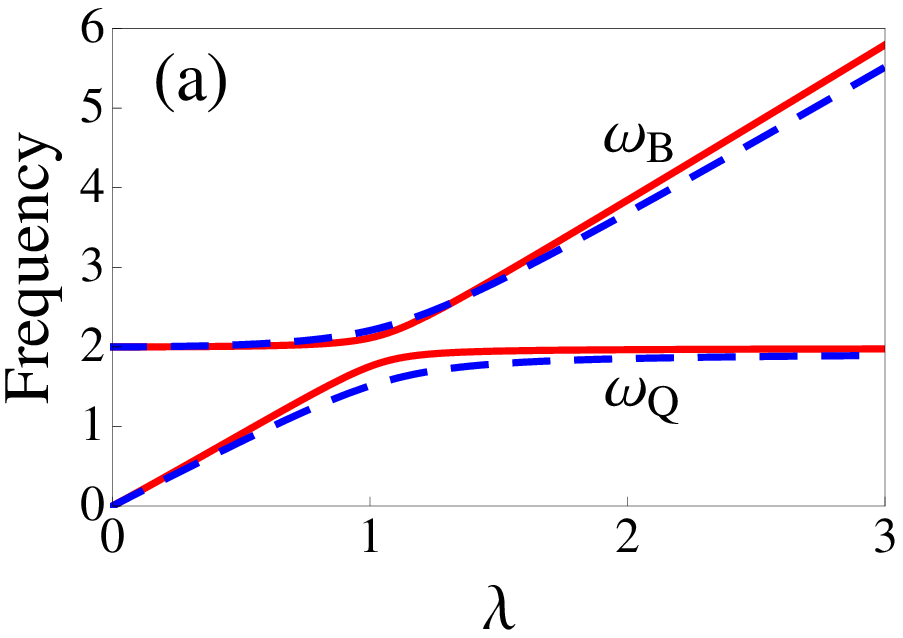}
\includegraphics[height=5cm]{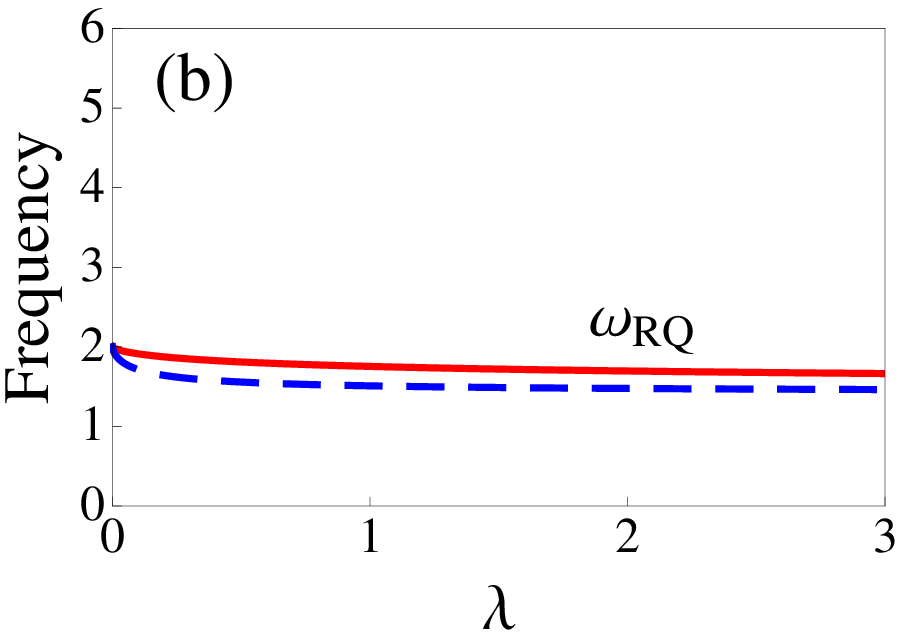}
\caption{Frequencies (in units of $\omega_\rho$) of collective oscillation modes for (a) breathing and quadrupole mode and (b) radial quadrupole mode versus trap aspect
ratio $\lambda$ for $p=1$, $k=0.001$ (solid red lines) and $p=10$, $k=0.1$ (dashed blue lines).}
\label{fig:frequencies}
\end{figure}

Using the Gaussian approximation enables us to analytically estimate frequencies
of the low-lying collective modes \cite{5c,8,9g}. This is
done by linearizing equations~(\ref{eq:ur}) and (\ref{eq:uz}) around the equilibrium
positions. If we expand the condensate widths as $u_{\rho}(t)=u_{\rho 0}+\delta
u_{\rho}(t)$ and $u_z(t)=u_{z0}+\delta u_z(t)$, insert these expressions into
the corresponding equations, and expand them around the equilibrium widths by keeping
only linear terms,
we immediately get the frequencies of the breathing and the quadrupole mode,
\begin{equation}
\omega_{\rm B,Q}^2 =\frac{m_1+m_3 \pm \sqrt{(m_1-m_3)^2+8m_2^2}}{2}\, ,
\label{eq:frequencyBQ}
\end{equation}
where the abbreviations $m_1, m_2$, and $m_3$ are given by
\begin{equation}
\hspace*{-5mm}
m_1= 4 +\frac{2k}{u_{\rho 0}^6 u_{z0}^2}\, , \quad
m_2=\frac{p}{u_{\rho 0}^3 u_{z0}^2}+\frac{2k}{u_{\rho 0}^5 u_{z0}^3}\, ,\quad
m_3=4\lambda^2 -\frac{ p}{u_{\rho 0}^2 u_{z0}^3}\, ,
\label{abb}
\end{equation}
and the corresponding breathing and quadrupole mode eigenvectors are given by
\begin{equation}
{\bf u}_{\rm B,Q}=\frac{1}{\sqrt{m_2^2+\left(\omega_{\rm B,Q}^2-m_1 \right) ^2
}} \
\left(\begin{array}{c} m_2\\ \omega_{\rm B,Q}^2-m_1
\end{array}\right)\, .
\label{eq:vector}
\end{equation}
The quadrupole mode has a lower frequency and is characterised by out-of phase
radial and axial oscillations, while in-phase oscillations correspond to the
breathing
mode. Another low-lying collective excitation is the radial quadrupole mode, which is characterised by out-of-phase oscillations in the $x$ and $y$ directions, while in the $z$ direction there are no oscillations. As this mode breaks the cylindrical symmetry, it can only be calculated by using the three-dimensional equations of motion. The frequency turns out to be
\begin{equation}
\omega_{\rm RQ}^2=2+\frac{2}{u_{\rho 0}^4}\, ,
\label{eq:frequencyRQ}
\end{equation}
and the corresponding three-dimensional eigenvector is
\begin{equation}
{\bf u}_{\rm RQ}=\frac{1}{\sqrt{2}} \,
\left(\begin{array}{c} 1\\ -1\\ 0\end{array}\right)\, .
\label{eq:vectorRQ}
\end{equation}
Figure~\ref{fig:frequencies} shows the frequencies of all collective oscillation modes as
functions of the trap aspect ratio $\lambda$.
We see that the collective mode frequencies depend relatively strongly on the trap
anisotropy, whereas a variation of the dimensionless
interaction strengths $p$ and $k$ yields only marginal changes.

\section{Stability Diagram}
\label{sec:SD}

In this section we discuss the stability of a Bose-Einstein condensate in the
mean-field framework for systems with two- and three-body contact interaction
in an axially-symmetric harmonic trap. It is well known that BEC systems with
an attractive two-body interaction are unstable against collapse above the critical
number of atoms (i.e. for sufficiently large negative value of $p$) in the condensate \cite{pitstri-book,petsmi-book}.
For smaller numbers of atoms, the zero-point kinetic energy is able
to counter the attractive inter-atomic interactions, however, when the number of atoms
sufficiently increases, this is no longer possible, and the system collapses to the centre
of the trapping potential.

We find that, for a pure two-body interaction, the condensate is stable only above a critical stability line $p_c(\lambda)$, while the presence of even a small repulsive three-body interaction leads to stabilisation of the condensate. On the other hand, we find that an attractive three-body interaction further destabilises the condensate.

To study in detail effects of three-body interaction on the stability of BEC systems, we consider
several cases of interest: repulsive and attractive pure two-body interactions,
attractive two-body and
repulsive three-body interactions, and attractive two- and three-body
interactions. If the corresponding system of equations does (not) have positive and bounded solutions of equations~(\ref{eq:ur}) and (\ref{eq:uz}) in the vicinity of
positive equilibrium widths determined by equations~(\ref{eq:ur0}) and (\ref{eq:uz0}), then the condensate is considered stable (unstable). This is equivalent to performing a linear stability
analysis and determining the stability of positive equilibrium widths by examining frequencies of the corresponding collective oscillation modes~(\ref{eq:frequencyBQ}) and (\ref{eq:frequencyRQ}). The solution is only stable if frequencies of all low-lying collective modes are found to be real, otherwise the solution is unstable.

\begin{figure}[!t]
\centering
\includegraphics[height=5cm]{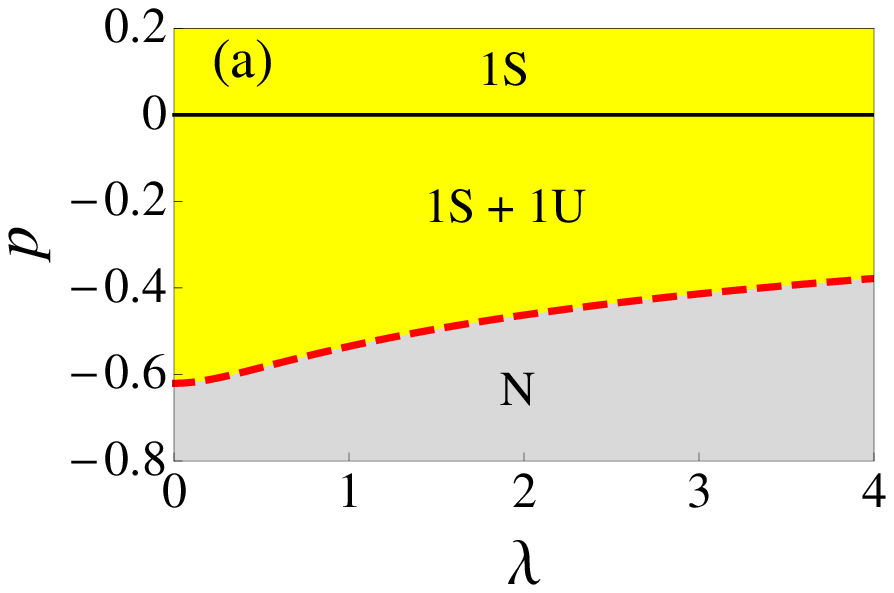}
\includegraphics[height=5cm]{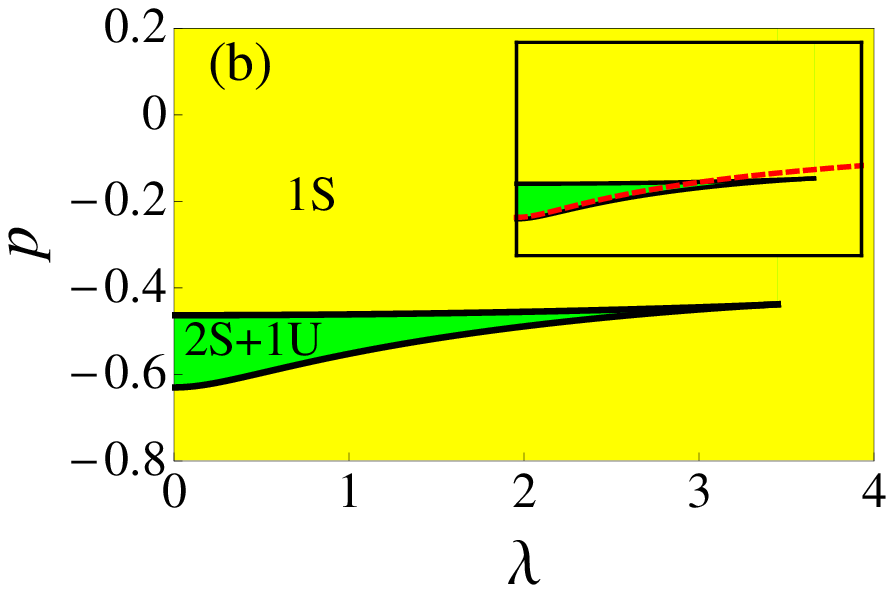}
\includegraphics[height=5cm]{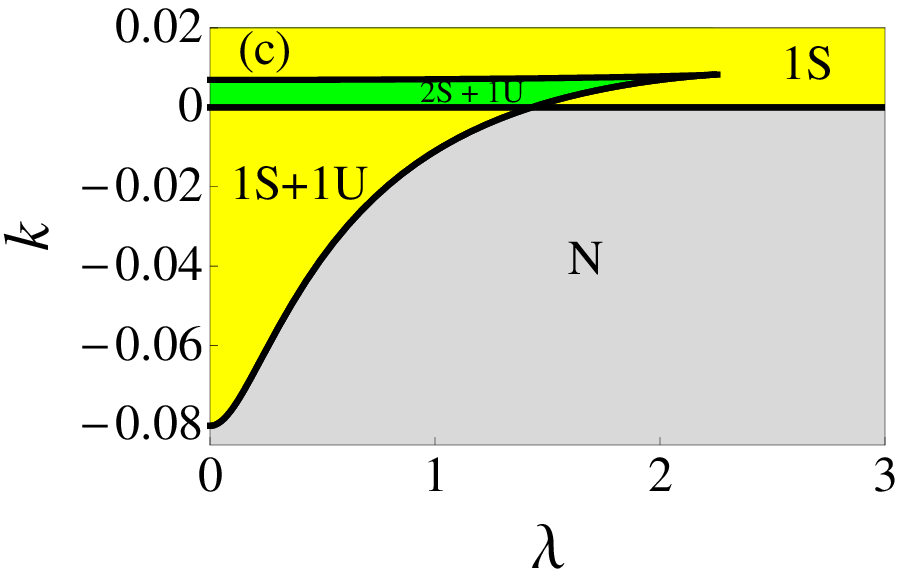}
\caption{Stability diagram of a BEC as a function of trap aspect ratio
$\lambda$ for different values of dimensionless two-body and three-body contact interaction strengths $p$ and $k$.
(a) $\lambda-p$ stability diagram for $k=0$, where the dashed red line represents the critical stability line, below which there are no solutions (N). Above this line, for $p<0$, there is one stable and one unstable solution (1S+1U), while for $p\geq 0$ there is only one stable solution (1S). (b) $\lambda-p$ stability diagram for $k=0.005$, where two cases exist: the small region with two stable and one unstable solution (2S+1U), while otherwise only one stable solution exists (1S). For comparison, in the inset we combine the critical stability line for $k=0$ with the stability diagram for $k=0.005$. (c) $\lambda-k$ stability diagram for $p=-0.5$. For $k\leq 0$, there are two regions: the one without solutions (N), and the one with one stable and one unstable solution (1S+1U). For $k> 0$, there are also two regions: the small region with two stable and one unstable solution (2S+1U), while otherwise there is only one stable solution (1S). As we can see, a non-vanishing value of the three-body interaction $k$ substantially enhances the stability of a condensate.}
\label{fig:SD1}
\end{figure}

For the case of
a pure repulsive two-body interactions, we will
immediately see that the condensate is always stable. For the case of
an attractive two-body interaction, the situation is quite different: the
above system of equations can have no equilibrium, or it could have
up to three equilibrium solutions. The
results of a detailed numerical
analysis are summarised in \fref{fig:SD1}.

\begin{figure}[!t]
\centering
\includegraphics[height=5cm]{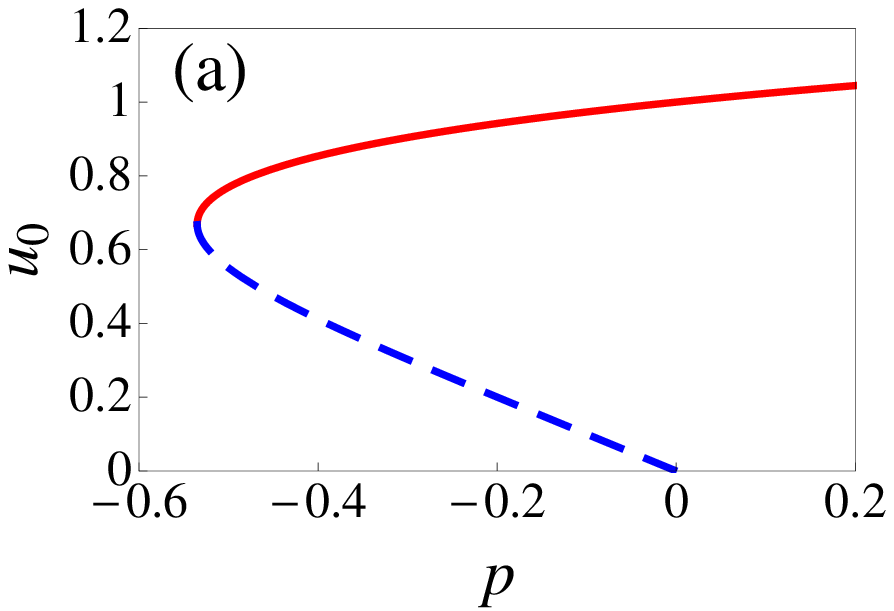}
\includegraphics[height=5cm]{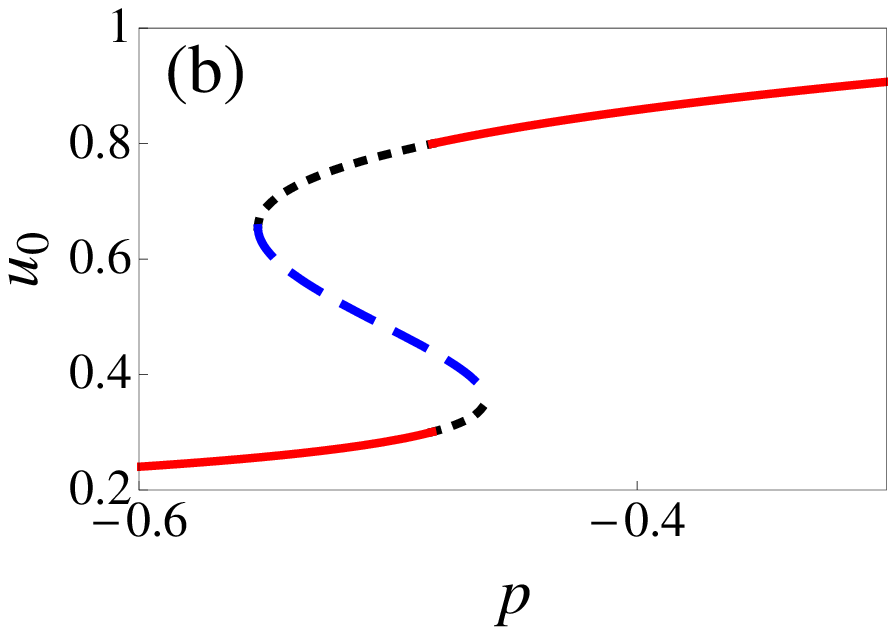}
\includegraphics[height=5cm]{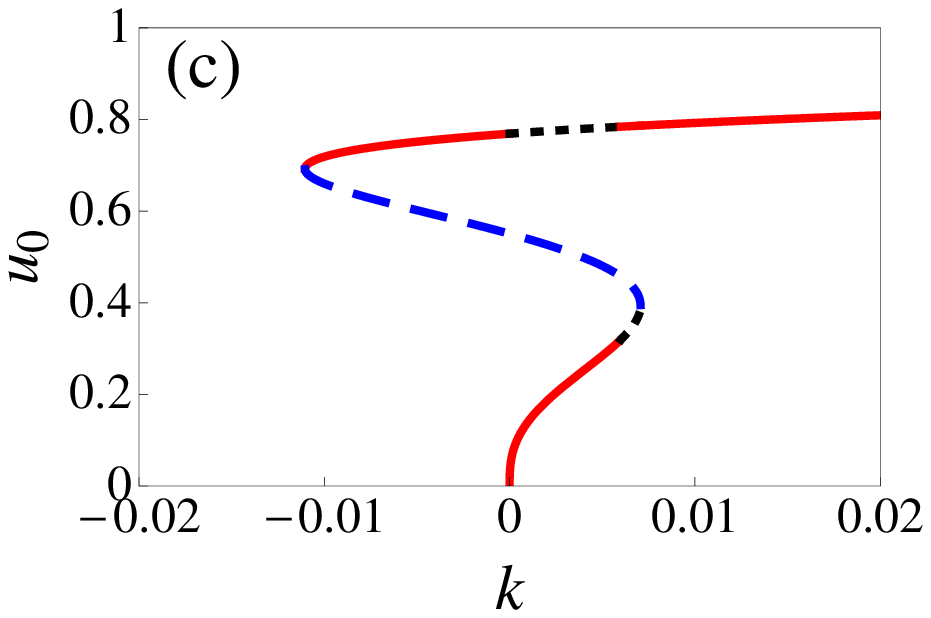}
\caption{Condensate width $u_{\rho 0}=u_{z0}=u_0$ for $\lambda=1$ and (a) $k=0$, as a function of $p$; (b) $k=0.005$, as a function of $p$; (c) $p=-0.5$, as a function of $k$.
Solid red lines represent the stable solution with minimal energy, dotted black lines represent another stable solution, dashed blue lines represent the unstable solution.}
\label{fig:u0}
\end{figure}

The dashed red line in \fref{fig:SD1}(a) represents the critical stability line as a
function of the trap aspect ratio $\lambda$ for a pure two-body interaction ($k=0$). Below the critical stability line there are no stable solutions, and the system is unstable. Above the critical stability line, the system has one stable and one unstable
solution for an attractive two-body interaction ($p<0$), and only one stable solution for a repulsive two-body interaction ($p\geq 0$).
For $\lambda =0$, which corresponds to the limit of a cigar-shaped
condensate, we have the critical value of two-body interactions $p_c=-0.6204$,
which coincides precisely with the value from reference~\cite{8}. For the isotropic case, when
$\lambda=1$, the critical
value is $p_c=-0.535$, which again coincides with the value from
the literature ~\cite{bec6,8,11e,11f}. \Fref{fig:u0}(a)~shows solutions for the isotropic condensate as a function of $p$.

Now, if we consider the case of an attractive two-body interaction and a small
repulsive three-body interaction, the results of the stability analysis are quite different. The system can either have one or three solutions, as shown in \fref{fig:SD1}(b). The presence of a positive three-body interaction $k$, however small, leads to the existence of at least one stable solution in the whole range of values of $\lambda$ and $p$. In the small area designated by 2S+1U in \fref{fig:SD1}(b), two stable and one unstable solution exist. Out of these two stable solutions, only the one with the minimal energy is physically relevant and could be realised in an experiment. \Fref{fig:u0}(b)~shows solutions for $\lambda=1$, $k=0.005$ as a function of $p$. As we can see, a minimal-energy stable solution exists for any value of $p$. However, for large negative values of $p$ this solution tends to zero, which practically represents a collapsed condensate. Therefore, although within the given mathematical model the condensate is always stable, physically this is valid only up to a critical number of atoms, which has to be determined by considering in detail the corresponding condensate density. However, as we can see from \fref{fig:u0}(b), the dependence $u_0(p)$ for large negative values of $p$ is quite flat, which means that the stability region can be significantly extended in the presence of a small positive value of $k$ compared to the case of pure two-body interaction.

We also analyse the stability of a BEC system as a function of three-body interaction $k$. \Fref{fig:SD1}(c)~shows the corresponding stability diagram for an attractive two-body interaction $p=-0.5$. For a repulsive three-body interaction ($k>0$), as expected, we see a small region with two stable and one unstable solution (2S+1U), as well as a region with only one stable solution (1S), similar to \fref{fig:SD1}(b). For an attractive three-body interaction ($k<0$), the stability region with one stable and one unstable solution (1S+1U), which corresponds to the 1S+1U region in \fref{fig:SD1}(a), gradually shrinks until it disappears as $k$ becomes sufficiently negative. Therefore, we see that an attractive three-body interaction has the same destabilising effect on a BEC as an attractive two-body interaction. This can be also seen in \fref{fig:u0}(c), where the stable minimal-energy solution for $p=-0.5$ exists only for a limited range of negative values of $k$.

\begin{figure}[!t]
\centering
\includegraphics[height=5cm]{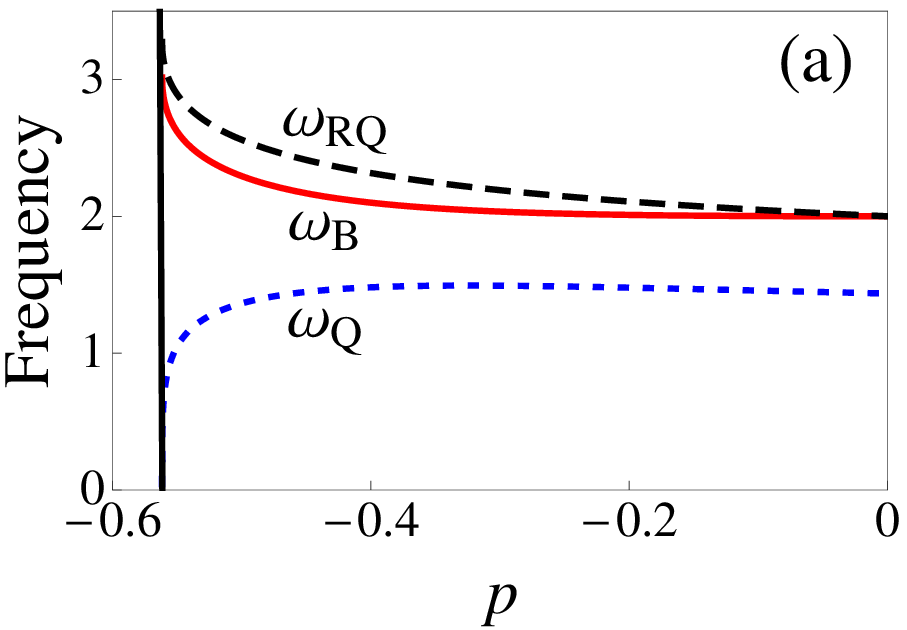}
\includegraphics[height=5cm]{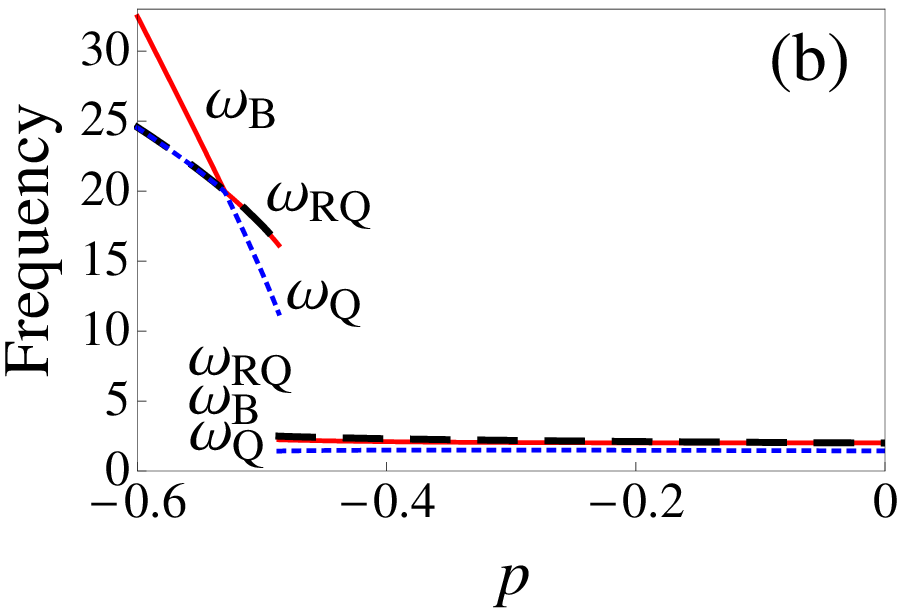}
\caption{Frequencies (in units of $\omega_\rho$) of low-lying collective excitation modes: breathing (B), radial
quadrupole (RQ), and quadrupole (Q), as functions of an attractive
two-body interaction $p$ for the trap anisotropy $\lambda=117/163$ and (a) $k=0$,
(b) $k=0.005$.}
\label{fig:freq}
\end{figure}

To further illustrate the findings of the above stability analysis, we plot in \fref{fig:freq}
the frequencies of the low-lying collective excitation modes as functions of an
attractive two-body interaction for the trap anisotropy $\lambda=117/163$
\cite{bec3}. \Fref{fig:freq}(a)~corresponds to the case when
three-body interactions are neglected, i.e. $k=0$, and we can see that the
condensate collapses for $p_c=-0.561$, when the expression for $\omega^2_{\rm Q}$ from equation~(\ref{eq:frequencyBQ}) becomes negative.
For a small repulsive three-body interaction $k=0.005$, \fref{fig:freq}(b)~shows the frequencies corresponding to stable minimum-energy solutions.
From \fref{fig:u0}(b) we see that for $p_c=-0.486$ there is a jump from one to another solution branch due to the minimal energy condition, which is reflected in \fref{fig:freq}(b) by
a corresponding jump in the frequencies of the collective modes.

\section{Shifts in frequencies of collective modes}
\label{sec:shifts}

Close to geometric resonances, the nonlinear structure of the GP equation (\ref{eq:gp}) leads to shifts
in the frequencies of collective oscillation modes compared to the respective
values in
equation~(\ref{eq:frequencyBQ}), which are calculated using a linear stability analysis. Here we
apply the standard Poincar\'e-Lindstedt method \cite{14,14a,14b,14c,9g} in order to
develop a perturbation theory and calculate these frequency shifts.

\subsection{Quadrupole mode}
\label{subsec:QM}

We start with working out a perturbation theory for the BEC dynamic, which is based on the set of ordinary
differential equations  (\ref{eq:ur})--(\ref{eq:uz}), by expanding the condensate widths
in the series
\begin{eqnarray}
\hspace*{-5mm}
u_{\rho}(t)&=&u_{\rho 0}+ \varepsilon u_{\rho1}(t) + \varepsilon^2 u_{\rho2}(t)
+
\varepsilon^3
u_{\rho3}(t)+\ldots \, ,\label{eq:pertu}\\
\hspace*{-5mm}
 u_z(t)&=&u_{z0}+ \varepsilon u_{z1}(t) + \varepsilon^2 u_{z2}(t) +
\varepsilon^3
u_{z3}(t)+\ldots \, ,\label{eq:pertu1}
\end{eqnarray}
where the smallness parameter $\varepsilon$ stems from the respective initial conditions. Here we
study the system dynamics with the initial conditions in the form
\begin{equation}
\label{eq:quadrupoleinitial}
{\bf u}(0)={\bf u}_0+\varepsilon {\bf u}_Q\, ,\quad
{\bf\dot{u}}(0)={\bf{0}}\, ,
\end{equation}
when the system is close to the equilibrium position ${\bf u}_0$, and is perturbed in the
direction of the quadrupole oscillation mode eigenvector ${\bf u}_Q$, determined by equation~(\ref{eq:vector}).
By inserting expansions (\ref{eq:pertu}) and (\ref{eq:pertu1}) into
equations~(\ref{eq:ur})--(\ref{eq:uz}), we obtain the following system of
linear differential equations:
\begin{eqnarray}
\ddot u_{\rho n}(t) + m_1 u_{\rho n}(t) + m_2 u_{zn}(t) &=&\chi _{\rho n}(t)\,
,\label{eq:urn}\\
\ddot u_{zn}(t) + 2 m_2 u_{\rho n}(t) + m_3 u_{zn}(t) &=& \chi_{zn}(t)\,
,\label{eq:uzn}
\end{eqnarray}
where the index $n$ takes integer values $n=1,2,3,..$, and the quantities $m_1$, $m_2$,
and $m_3$ are already defined by expressions (\ref{abb}). The
functions $\chi_{\rho n}(t)$ and $\chi_{zn}(t)$ depend only on the solutions
$u_{\rho
i}(t)$ and $u_{zi}(t)$ of lower orders $i$, i.e. those corresponding to
$i<n$. Therefore, the above system of equations can be solved hierarchically, and at each level $n$ of this procedure we use the initial
conditions from equations~(\ref{eq:quadrupoleinitial}).

In order to decouple the system of equations (\ref{eq:urn})--(\ref{eq:uzn}), we use the linear
transformation
\begin{eqnarray}
u_{\rho n}(t)=x_n(t)+y_n(t)\, ,\\
u_{zn}(t)=c_1 x_n(t)+ c_2 y_n(t)\,
\end{eqnarray}
with the coefficients
\begin{equation}
\label{eq:c12}
c_{1,2}=\frac{m_3-m_1\mp \sqrt{(m_3-m_1)^2+ 8 m_2^2}}{2 m_2}\, ,
\end{equation}
which leads to two independent linear second-order differential equations:
\begin{eqnarray}
\ddot x_n(t)+ \omega_Q^2 x_n(t)+\frac{c_2\chi_{\rho
n}(t)-\chi_{zn}(t)}{c_1-c_2}=0\, ,\\
\ddot y_n(t)+ \omega_B^2 y_n(t)+\frac{\chi_{zn}(t)-c_1 \chi_{\rho
n}(t)}{c_1-c_2}=0\, .
\end{eqnarray}
From this we see that $x_n(t)$
and $y_n(t)$ correspond to quadrupole and breathing mode
oscillations, respectively. Although the system is initially perturbed only in the direction of the
quadrupole mode eigenvector, due to the nonlinearity of the system, the breathing
mode is excited as well.
The solutions of the above equations depend essentially on the nature of
the inhomogeneous terms, which are given by polynomials of harmonic functions of
$\omega_Q t$, $\omega_B t$ and their linear combinations $(k\omega_Q
+m\omega_B)t$. Therefore, compared to linear systems, the important difference here is
that higher harmonics and linear combinations of the modes emerge due to the structure of GP equation.

A careful analysis also reveals the important conclusion that secular terms will start appearing at the level $n=3$.
As usual, they can be absorbed by a shift in the quadrupole mode frequency \cite{9g,14,14a,14b,14c}.
At level $n=3$, equations~(\ref{eq:urn})--(\ref{eq:uzn}) can be written as
\begin{equation}
{\bf{\ddot u}}_3(t)+{\cal M} {\bf u}_3(t) + {\bf I}_{Q,3} \cos \omega_Q t +\ldots=0\, ,
\label{eq:3order}
\end{equation}
with the matrix $\cal M$ defined as
\begin{equation}
{\cal M}=\left(\begin{array}{cc} m_1 & m_2\\ 2m_2 & m_3 \end{array}\right)\, ,
\end{equation}
and the dots represent the inhomogeneous part of the equation, which does not contain linear terms
proportional to harmonic functions in $\omega_Q t$. The expression for ${\bf
I}_{Q,n}$ can be calculated systematically in {\it Mathematica} software package \cite{Mathematica}.

The frequency shift of the quadrupole mode is found to be quadratic in $\varepsilon$:
\begin{equation}
 \omega_Q(\varepsilon) = \omega_Q+ \Delta \omega_Q =\omega_Q-\varepsilon^2
\frac{({\bf u}_Q^L)^T {\bf I}_{Q,3}}{2 \omega_Q }\, ,
\label{eq:quadrupoleshift}
\end{equation}
where ${\bf u}_Q^L$ is the left-hand quadrupole mode eigenvector of the matrix $\cal M$.
After a detailed calculation, the frequency shift of a quadrupole mode to lowest order
in $\varepsilon$ is found to be
\begin{equation}
\Delta \omega_Q= -\varepsilon^2\frac{f_{Q,3}(\omega_Q, \omega_B, u_{\rho 0},
u_{z0}, p,k,
\lambda)}{2\omega_Q(\omega_B-2\omega_Q)(\omega_B+ 2\omega_Q)}\, ,\label{eq:fsQ}
\end{equation}
where $f_{Q,3}$ is a regular function, without poles for real values of its
arguments.
The above expression (\ref{eq:fsQ}) has a pole for
$\omega_B=2\omega_Q$. Taking into account fact that $\omega_Q<\omega_B$, as we can
see from equation~(\ref{eq:frequencyBQ}) and \fref{fig:frequencies}, as well as the fact that collective frequencies depend on the trap aspect ratio $\lambda$, the condition $\omega_B=2\omega_Q$ can, in principle, be satisfied.
This is denoted as a geometric resonance, since it is obtained by simply tuning the geometry of the experiment through $\lambda$.
Higher-order corrections to $\Delta \omega_Q$ in $\varepsilon$
could, in principle, be obtained systematically by using the developed perturbation theory.

\subsection{Breathing mode}
\label{subsec:BM}

In a similar manner, we also study the dynamics of a cylindrically-symmetric BEC system when
initially only the breathing mode is excited,
\begin{equation}
\label{eq:breathinginitial}
{\bf u}(0)={\bf u}_0+\varepsilon {\bf u}_B\, ,\quad
{\bf\dot{u}}(0)={\bf{0}}\, .
\end{equation}
Applying again the Poincar\'e-Lindstedt perturbation theory, we calculate the
breathing mode frequency shift,
\begin{equation}
 \omega_B(\varepsilon) = \omega_B + \Delta \omega_B =\omega_B-\varepsilon^2
\frac{({\bf
u}_B^L)^T {\bf I}_{B,3}}{2 \omega_B }\, ,
\label{eq:breathingshift}
\end{equation}
where again the expression $({\bf u}_B^L)^T {\bf I}_{B,3}$ is calculated in {\it Mathematica}.
In this way we finally yield the following analytic formula for the frequency shift of the
breathing mode
\begin{equation}
\Delta \omega_B= -\varepsilon^2\frac{f_{B,3}(\omega_Q, \omega_B, u_{\rho 0},
u_{z0}, p,k,
\lambda)}{2\omega_B(2\omega_B-\omega_Q)(2\omega_B+ \omega_Q)}\, ,\label{eq:fsB}
\end{equation}
where the function $f_{B,3}$ is a regular function of its arguments. Naively looking
at this expression, one would say that it exhibits a pole for
$2\omega_B=\omega_Q$. However, from equation~(\ref{eq:frequencyBQ}) and \fref{fig:frequencies} we see that
$\omega_Q<\omega_B$, and, therefore, the condition $2\omega_B=\omega_Q$ is never
satisfied. In the next subsection we numerically demonstrate that a geometric resonance does not occur, and verify the
analytical result for the frequency shift of the breathing mode.

\subsection{Comparison with numerical results}
\label{subsec:numerics}

\begin{figure}[!t]
\centering
\includegraphics[height=4.8cm]{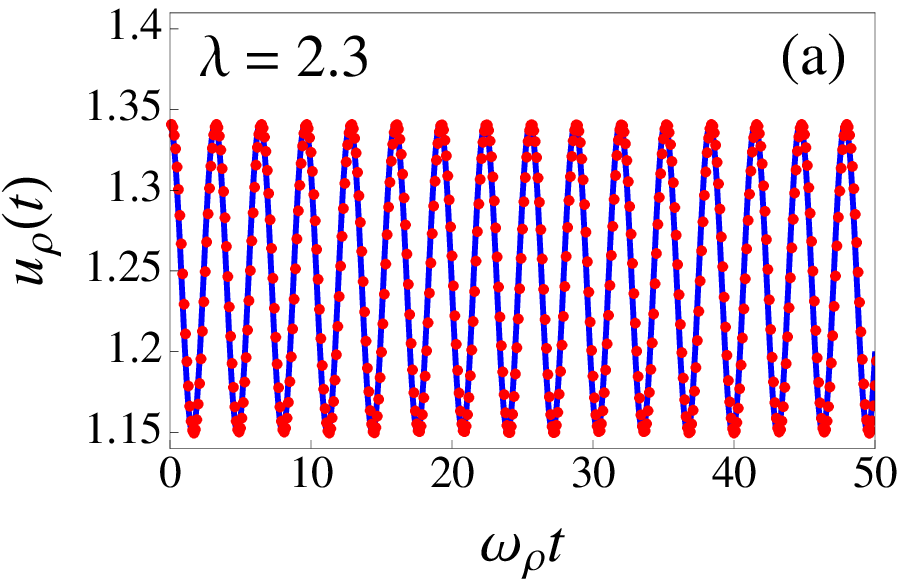}
\includegraphics[height=4.8cm]{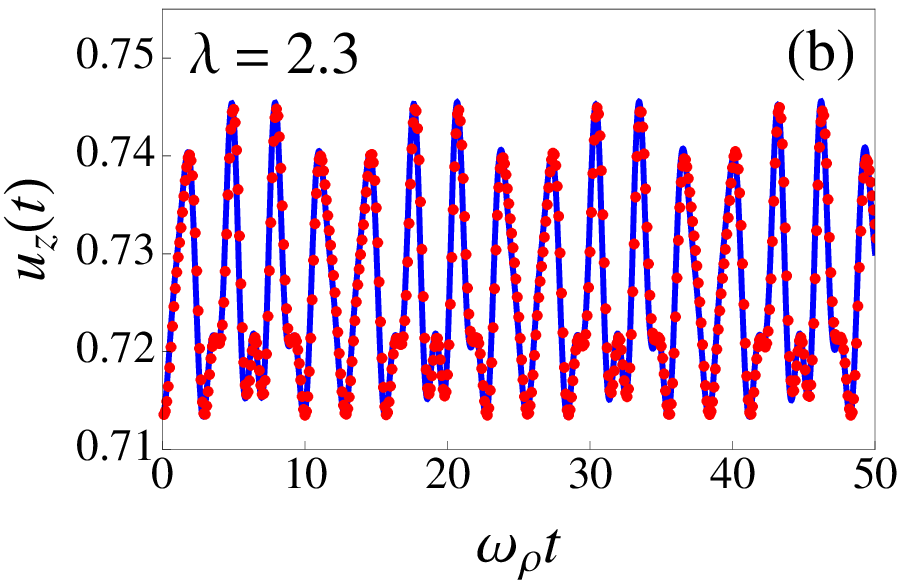}\\
\includegraphics[height=4.8cm]{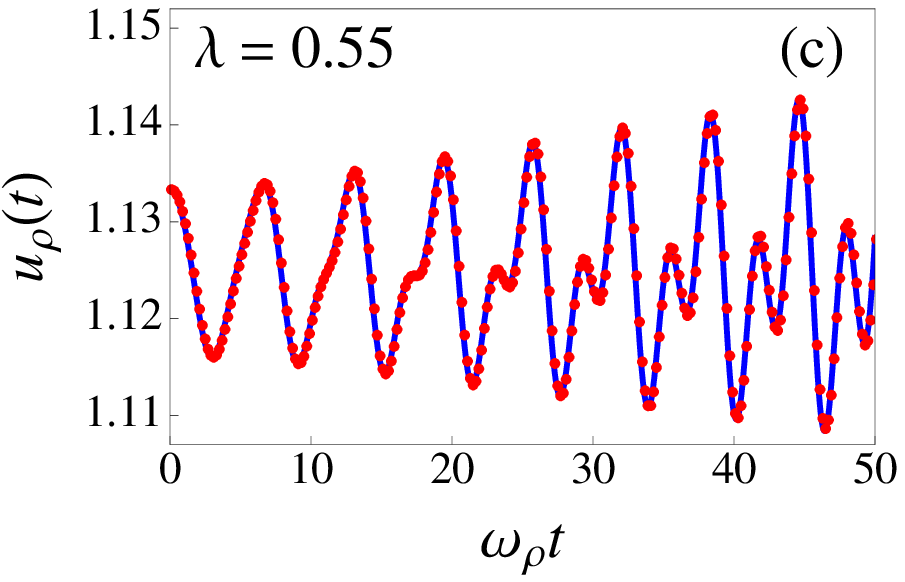}
\includegraphics[height=4.8cm]{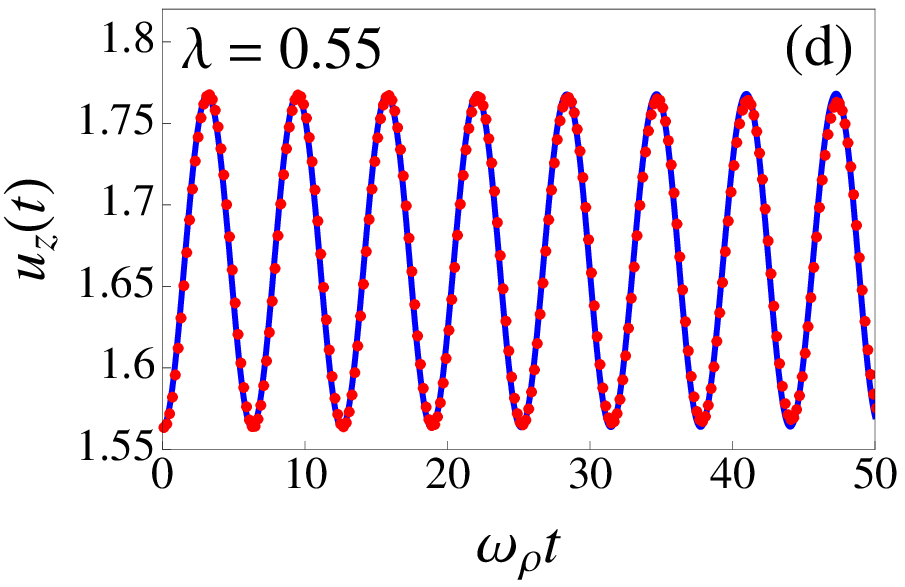}
\caption{A comparison of analytic (solid blue lines) and numeric (red dots)
results for a BEC dynamics with a pure repulsive two-body interaction
$p=1$, $k=0$, and $\varepsilon=0.1$. Top panels show dynamics of (a)
radial and (b) longitudinal condensate widths for the trap aspect ratio
$\lambda=2.3$ as a function of the dimensionless time $\omega_\rho t$;
bottom panels show dynamics of (c) radial and (d) longitudinal BEC widths
for $\lambda=0.55$.}
\label{fig:dc0}
\end{figure}

\begin{figure}[!t]
\centering
\includegraphics[height=4.8cm]{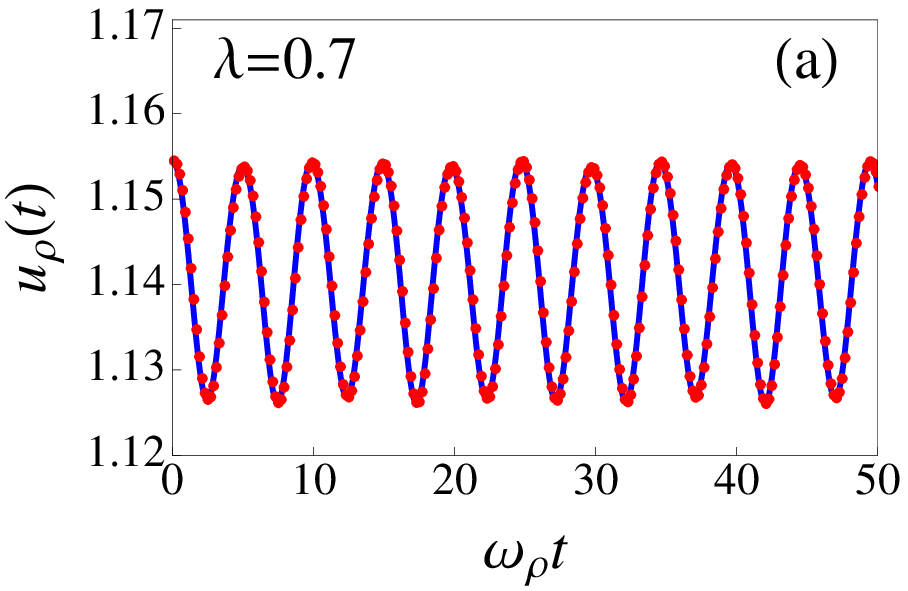}
\includegraphics[height=4.8cm]{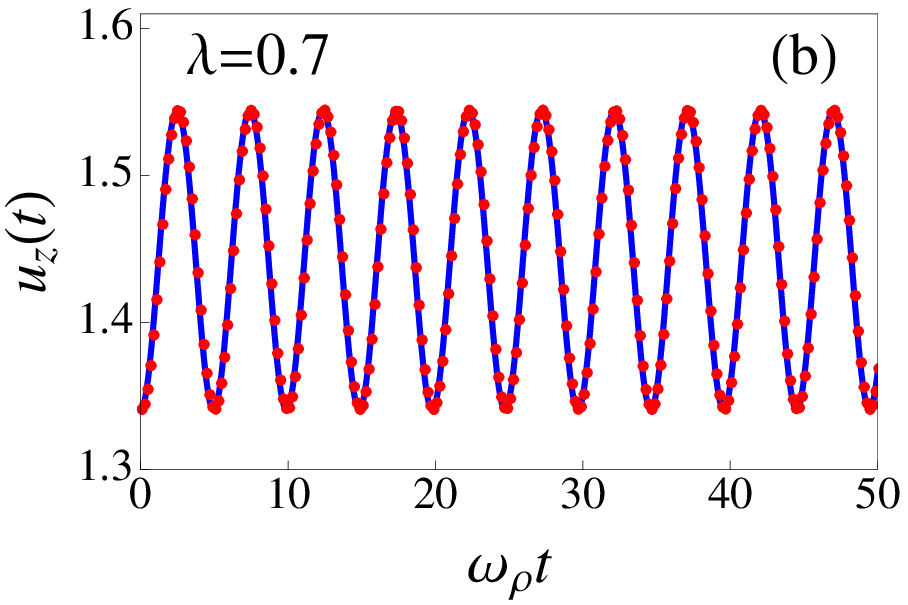}\\
\includegraphics[height=4.8cm]{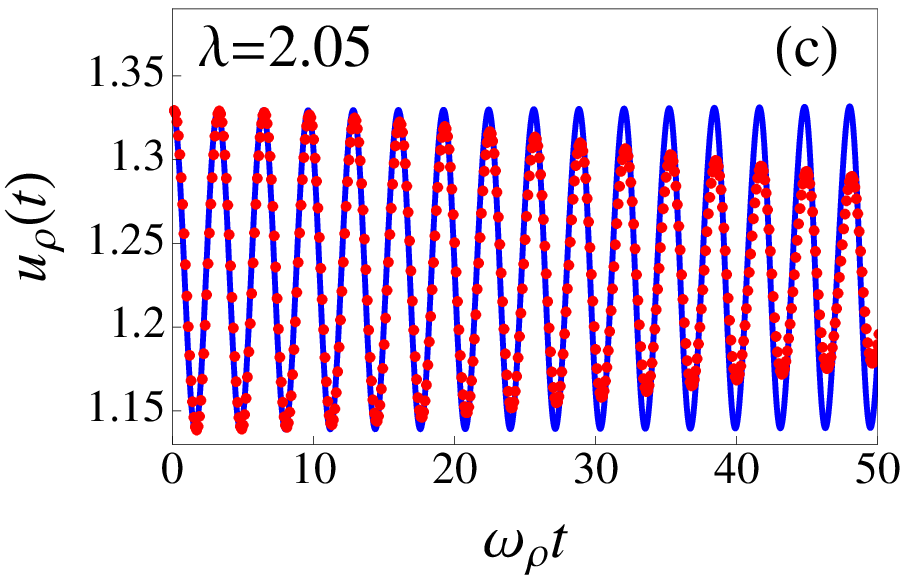}
\includegraphics[height=4.8cm]{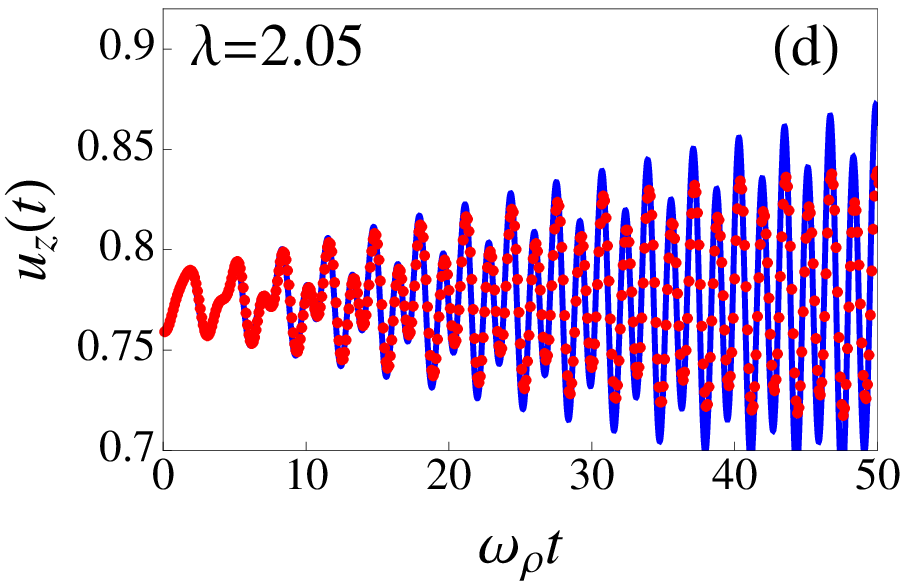}
\caption{A comparison of analytic (solid blue lines) and numeric (red dots)
results for BEC dynamics for a repulsive two-body interaction $p=1$
and a repulsive three-body interaction $k=0.001$, with $\varepsilon=0.1$. Top
panels show dynamics of (a) radial and (b) longitudinal condensate widths for
the trap aspect ratio $\lambda=0.7$ as a function of the dimensionless time
$\omega_\rho t$; bottom panels show dynamics of (c) radial and
(d) longitudinal BEC widths for $\lambda=2.05$.}
\label{fig:dc1}
\end{figure}

In order to verify our analytical results, we perform high-precision numerical
simulations \cite{GP-SCL1,GP-SCL2,abpp1,abpp2,abpp3,abpp4,abpp5,abpp6,abpp7}. At first we focus on a description of the BEC dynamics, and
compare our analytical results for the radial and longitudinal widths of the
condensate obtained perturbatively to the direct numerical solutions of
equations~(\ref{eq:ur})--(\ref{eq:uz}). To this end we consider a BEC in the initial state
corresponding to the perturbed equilibrium position, where the small perturbation is
proportional to the eigenvector of the quadrupole mode according to equations~(\ref{eq:quadrupoleinitial}). Examples of
the condensate dynamics are shown in \fref{fig:dc0} for a pure two-body
interaction $p = 1$, $k=0$ with $\varepsilon=0.1$, and in \fref{fig:dc1} for $p =
1$, $k=0.001$, $\varepsilon=0.1$.

In both figures we plot analytical and numerical solutions for $u_\rho$ and
$u_z$ as functions of the dimensionless time parameter $\omega_\rho t$ for
different values of the trap aspect ratio $\lambda$. Analytical solutions are
calculated using the third-order perturbation theory developed in
subsection~\ref{subsec:QM}. We can see in \fref{fig:dc0} that the agreement is
excellent, not only for the non-resonant value of the trap aspect ratio $\lambda=2.3$
(top panels), but also for $\lambda=0.55$ (bottom panels), which is close to a geometric resonance, as we will see later in \fref{fig:frequencyshiftquad}(a). 
For these values of parameters, the relative shift in the quadrupole mode frequency is of the order of 0.3\%, and therefore third-order perturbation theory
yields a quite accurate description of the system dynamics.
The same applies to the top panels of \fref{fig:dc1}, where $\lambda=0.7$ is far from any resonance. However, for $\lambda=2.05$ (bottom panels) we observe some disagreement, which increases
with propagation time.
This is due to the fact that $p=1$, $k=0.001$, $\lambda=2.05$ is close to a geometric resonance, as we will see in \fref{fig:frequencyshiftquad}(b).
In this case the perturbatively calculated shift in the quadrupole mode frequency is much larger than for the bottom panels of \fref{fig:dc0}. For this reason, after a long enough time the third-order perturbation theory is not sufficiently accurate. Although it gives a qualitatively correct description of the behaviour of the system, one would have to go to higher orders in perturbation
theory to get a more accurate agreement with the numerical results.
Such a behaviour in the bottom panels of \fref{fig:dc1} is just a tell-tale of the occurrence of a geometric resonance, and a subsequent analysis of frequency shifts is the only proper way to identify these resonances in a more quantitative way.

\begin{figure}[!t]
\centering
\includegraphics[height=5cm]{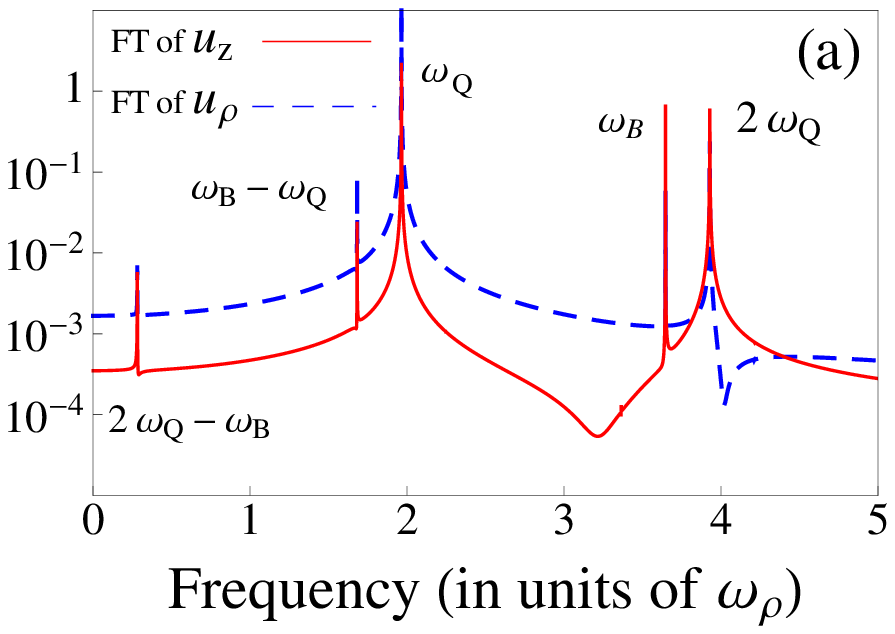}
\includegraphics[height=5cm]{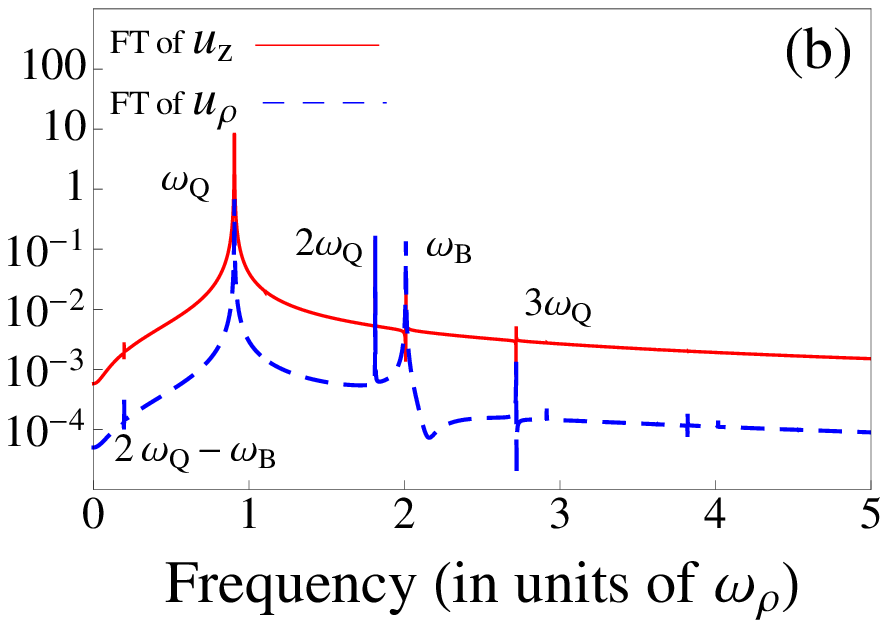}
\caption{Fourier spectra of the BEC dynamics obtained by numerically solving the system of equations~(\ref{eq:ur}) and (\ref{eq:uz}) for a repulsive two-body
interaction $p=1$, a repulsive three-body interaction $k=0.001$, and
$\varepsilon=0.1$ for (a) $\lambda=1.9$ and (b) $\lambda=0.5$. Each graph shows
spectra of both longitudinal and radial condensate widths. The locations of all peaks are
identified as linear combinations of the quadrupole and the breathing mode frequency,
in correspondence with the analysis based on the developed perturbation theory.}
\label{fig:excitation}
\end{figure}

However, before we present this analysis, we show in \fref{fig:excitation} the excitation
spectra of the BEC the dynamics which corresponds to the initial conditions
(\ref{eq:quadrupoleinitial}) for $p=1$,
$k=0.001$, and two values of the trap aspect ratio, $\lambda=1.9$ and
$\lambda=0.5$. For the parameter values in \fref{fig:excitation}(a),
the linear stability analysis yields breathing and
quadrupole mode frequencies (\ref{eq:frequencyBQ}) with $\omega_B=3.65$ and $\omega_Q=1.96$, while for
the parameters in \fref{fig:excitation}(b) we obtain $\omega_B=2.01$ and
$\omega_Q=0.905$, all expressed in units of $\omega_\rho$. In both graphs we
can see that the Fourier spectra contain two basic modes, $\omega_Q$ and $\omega_B$,
whose values agree well with those obtained from the linear stability analysis in equation~(\ref{eq:frequencyBQ}),
and a multitude of higher-order harmonics, which are linear combinations of the two modes,
as pointed out in subsection~\ref{subsec:QM}.

\begin{figure}[!t]
\centering
\includegraphics[height=4.5cm]{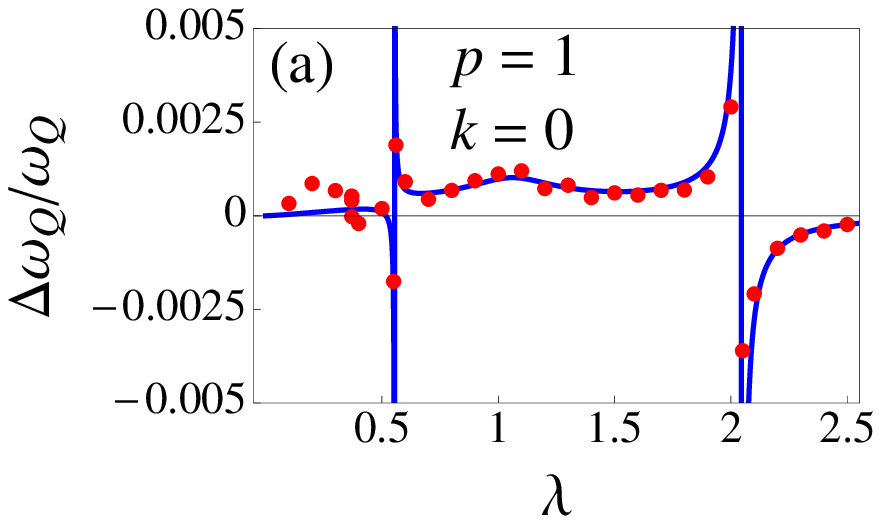}
\includegraphics[height=4.5cm]{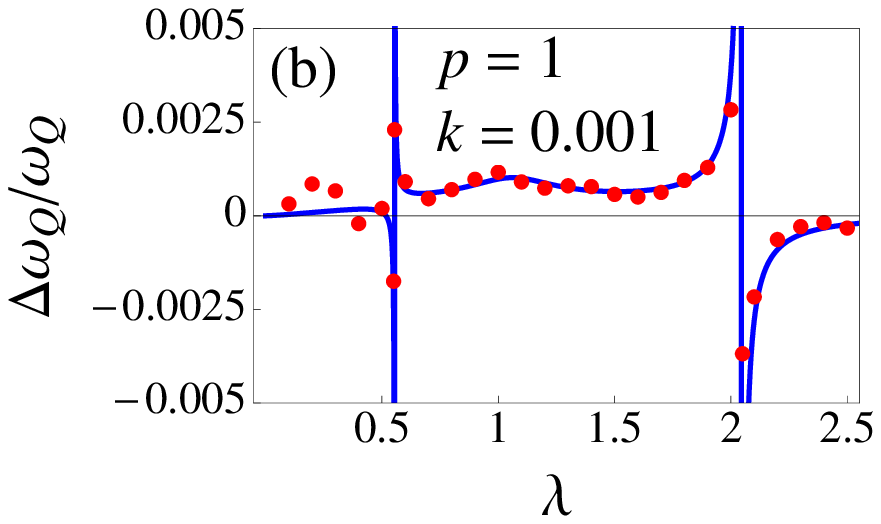}\\
\includegraphics[height=4.5cm]{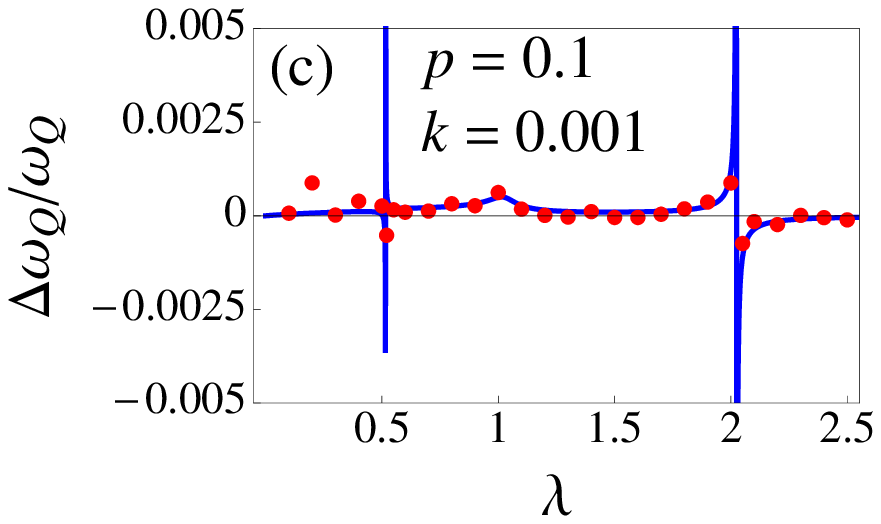}
\includegraphics[height=4.5cm]{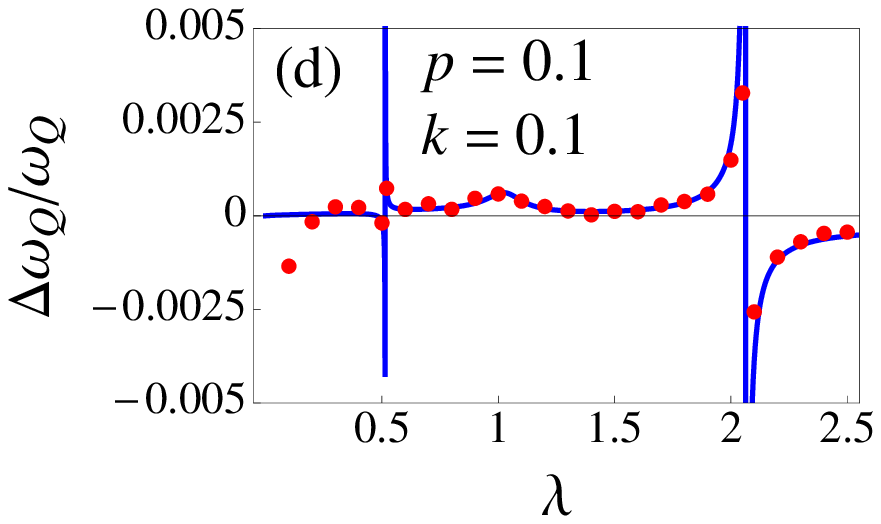}\\
\includegraphics[height=4.5cm]{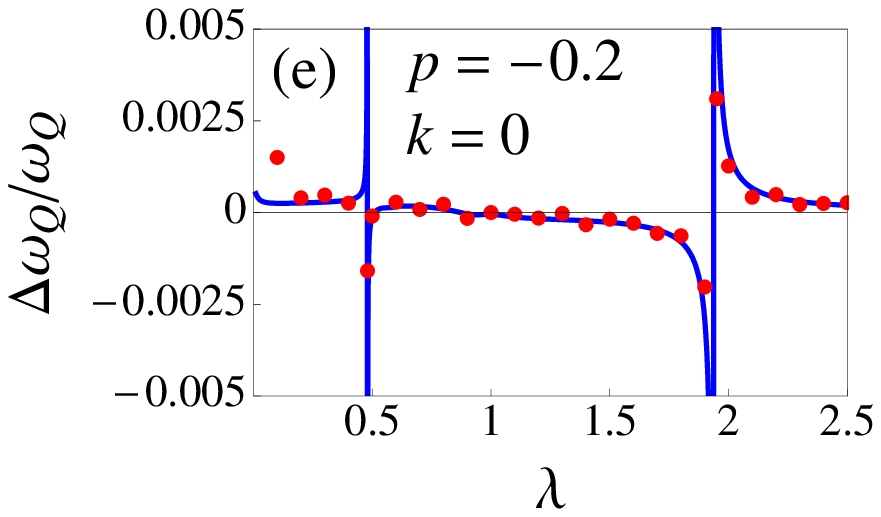}
\includegraphics[height=4.5cm]{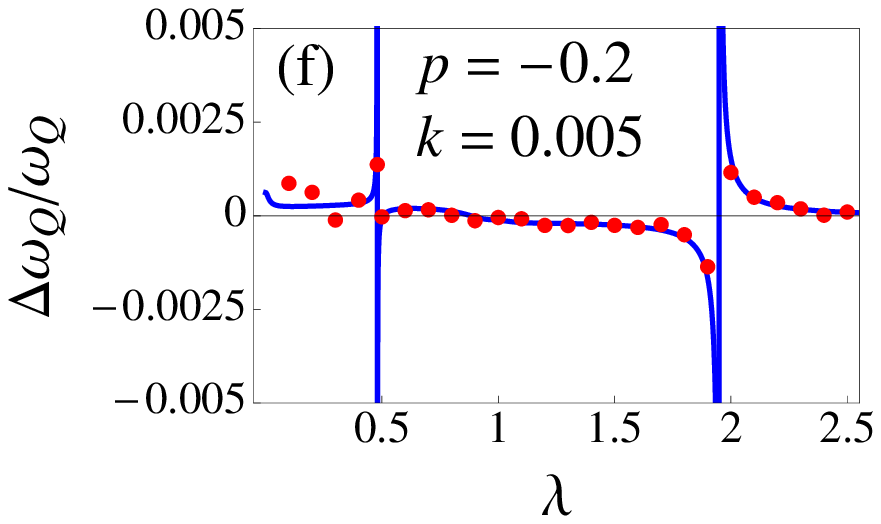}
\caption{Relative frequency shift of the quadrupole mode as a function of the
trap aspect
ratio $\lambda$ for $\varepsilon=0.1$ and different values of two-body and
three-body interaction strengths: (a) $p=1$, $k=0$, (b) $p=1$, $k=0.001$,
(c) $p=0.1$, $k=0.001$, (d) $p=0.1$, $k=0.1$, (e) $p=-0.2$, $k=0$, (f) $p=-0.2$,
$k=0.005$. Solid lines represent the analytical result
(\ref{eq:fsQ}), while dots are obtained by a numerical analysis of the
corresponding excitation spectrum for each value of $\lambda$, as described in
\fref{fig:excitation}.}
\label{fig:frequencyshiftquad}
\end{figure}

\begin{figure}[!t]
\centering
\includegraphics[height=4.5cm]{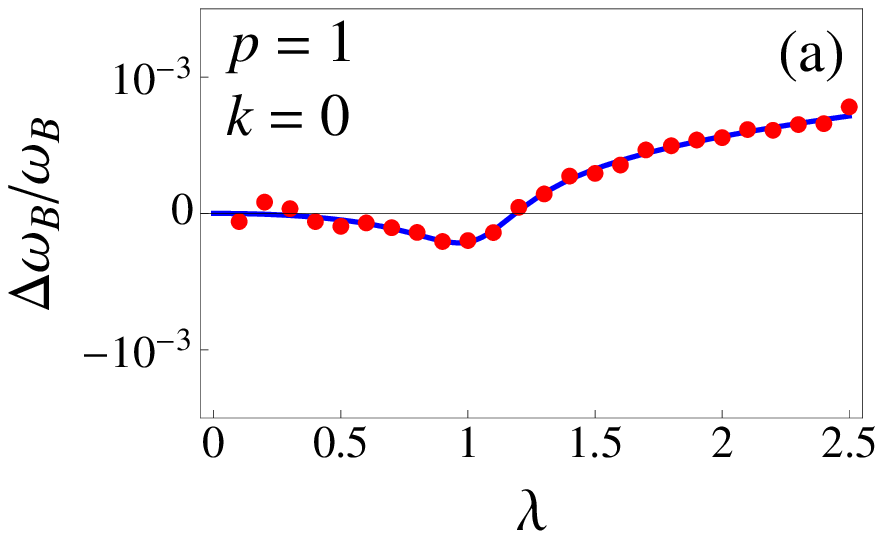}
\includegraphics[height=4.5cm]{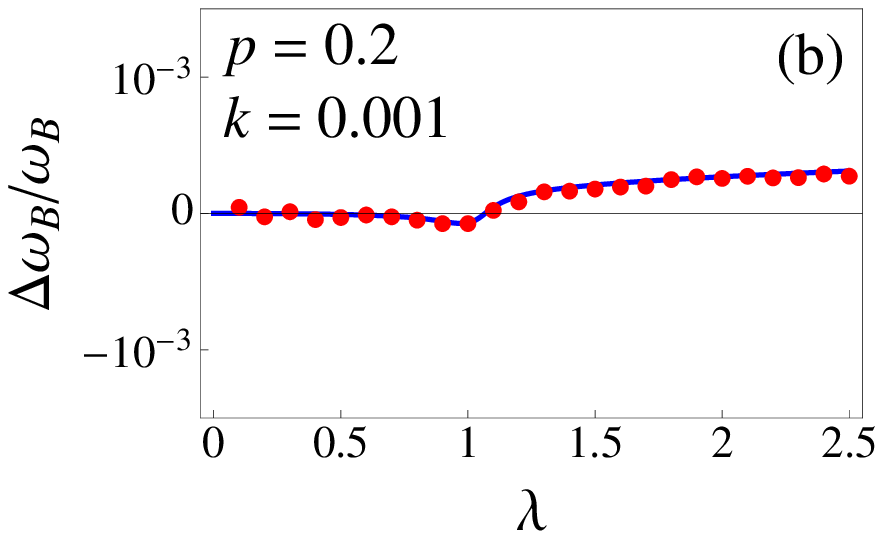}\\
\includegraphics[height=4.5cm]{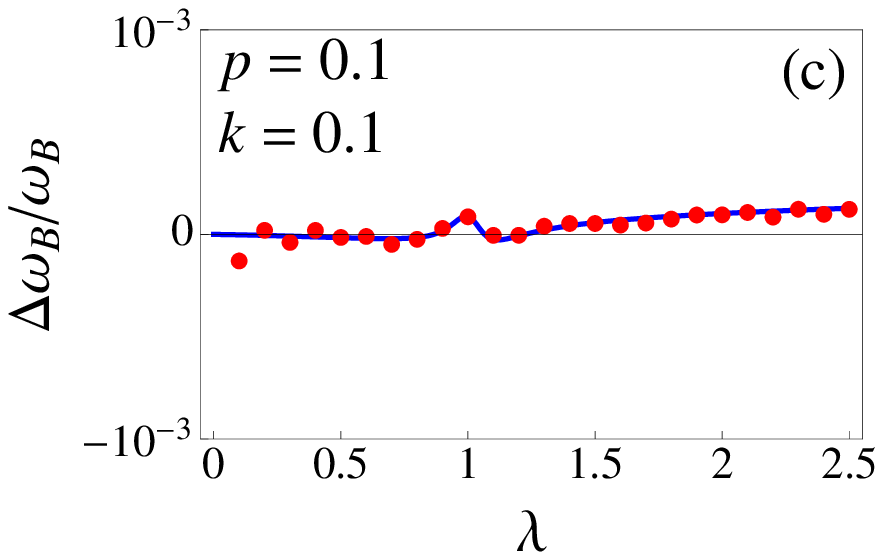}
\includegraphics[height=4.5cm]{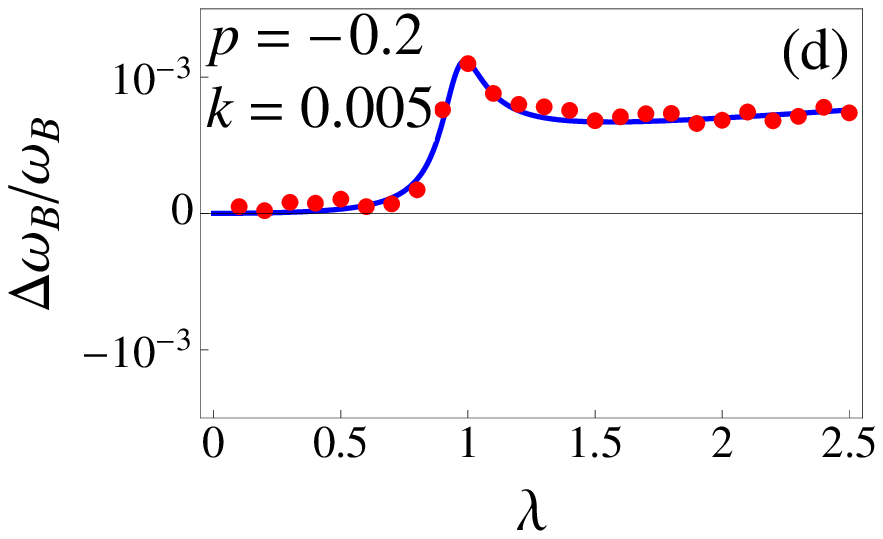}
\caption{Relative frequency shift of the breathing mode as a function of the
trap aspect
ratio $\lambda$ for $\varepsilon=0.1$ and different values of two-body and
three-body interaction strengths: (a) $p=1$, $k=0$, (b) $p=0.2$, $k=0.001$,
(c) $p=0.1$, $k=0.1$, (d) $p=-0.2$, $k=0.005$. Solid lines represent the analytical
result (\ref{eq:fsB}), while dots are obtained by a numerical
analysis of the corresponding excitation spectrum for each value of $\lambda$,
as described in \fref{fig:excitation}.}
\label{fig:frequencyshiftbre}
\end{figure}

Now we compare the derived analytical results for the frequency shifts of the
quadrupole and the breathing mode with the results of numerical simulations for
the BEC systems with two- and three-body contact interactions in a cylindrical trap.
In particular, we note that the calculated frequency shifts
close to geometric resonances reveal poles, which are an artefact of the perturbative approach.
Indeed, our detailed numerical calculations show that the observed frequencies remain finite through the whole geometric resonance.
In figures~\ref{fig:frequencyshiftquad} and \ref{fig:frequencyshiftbre} we
present the comparison of analytic (solid lines) and numeric (dots) values of
relative frequency shifts as functions of the trap aspect ratio $\lambda$. The
analytical results are calculated from equation~(\ref{eq:fsQ}) and (\ref{eq:fsB}), respectively, 
while the numerical data are obtained from a Fourier analysis of the excitation
spectrum, i.e. for each value of $\lambda$ we have calculated the corresponding
Fourier spectra, as in \fref{fig:excitation}, and then extracted the frequency values of the quadrupole and the breathing mode.

In \fref{fig:frequencyshiftquad}(a) we show a special case of a pure
two-body interaction, when $k=0$. The condition for a geometric resonance
$\omega_B=2\omega_Q$ yields the trap aspect ratios $\lambda_1 = 0.555$ and $\lambda_2 = 2.056$,
in good agreement with the numerical data, as we can see from the graph. The
existence of a geometric resonance at $\lambda_1=0.555$ is responsible for a
violent dynamics seen in the bottom panels of \fref{fig:dc0}, as we have
pointed out earlier. However, by analysing the frequency shifts we can
conclusively show that, indeed, the geometric resonance is present.
In further graphs we see that the excellent agreement between analytical and
numerical results holds also for other values of $p$ and $k$, including the case
of an attractive two-body interaction $p=-0.2$, which is still within the BEC
stability region. It is interesting to note the observation that the asymptotic approach to
geometric resonances for the case of an attractive two-body interaction is reversed
compared to the case of a repulsive two-body interaction. For instance, we can
see in \fref{fig:frequencyshiftquad}(d) that $\Delta
\omega_Q/\omega_Q\to\infty$ when $\lambda\to\lambda_2^-$, and $\Delta
\omega_Q/\omega_Q\to-\infty$ when $\lambda\to\lambda_2^+$, while for an
attractive $p=-0.2$ in \fref{fig:frequencyshiftquad}(f) we see that the
situation is reversed.

In \fref{fig:frequencyshiftbre} we compare analytic and numeric results for
a frequency shift of the breathing mode. As for the quadrupole mode, the
agreement is excellent for both repulsive and attractive two-body interaction.
As pointed out in subsection~\ref{subsec:BM}, there are no geometric resonances for
the breathing mode frequency, since the corresponding condition
$\omega_Q=2\omega_B$ cannot be satisfied.

\begin{figure}[!t]
\centering
\includegraphics[height=4.5cm]{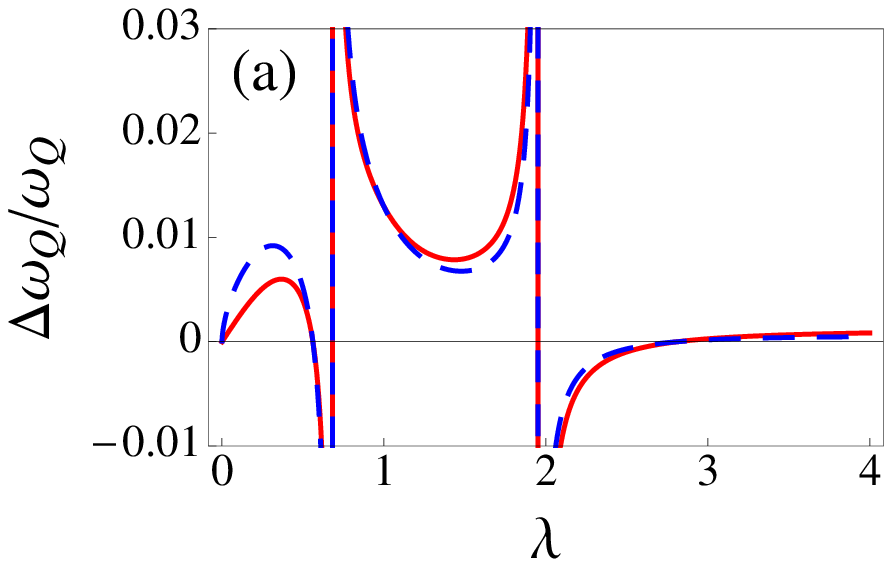}
\includegraphics[height=4.5cm]{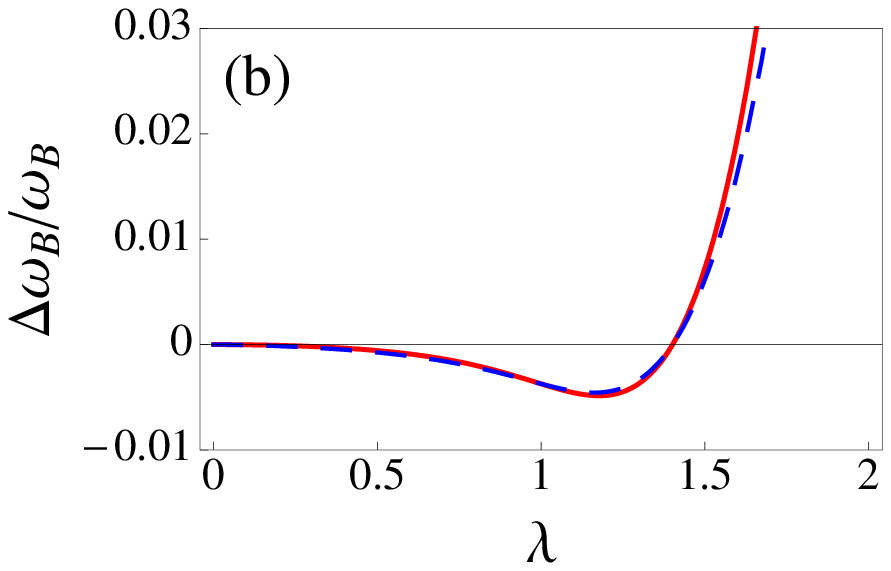}
\caption{Comparison of the analytical results for the relative frequency shifts of
(a) quadrupole and (b) breathing mode in the Thomas-Fermi limit from
reference~\cite{c9} derived using the parabolic variational ansatz (solid red lines) and the analytical results derived here using the
Poincar\'e-Lindstedt perturbation theory with the Gaussian variational ansatz (dashed blue lines).}
\label{fig:comparison}
\end{figure}

Finally, we compare our derived analytic results with those from reference~\cite{c9},
where frequency shifts of collective modes were calculated in the Thomas-Fermi
(TF) limit using a hydrodynamic approach. In terms of our variational approach, the
TF limit corresponds to the limit $p\rightarrow \infty$, so that
equation~(\ref{eq:frequencyBQ}) for the frequencies of the breathing and the quadrupole
mode reduces to
\begin{equation}
 \omega^2_{B,Q}= 2 + \frac{3}{2} \lambda^2 \pm \frac{1}{2}\sqrt{16 - 16\lambda^2
+ 9 \lambda^4 }\, ,
\end{equation}
in agreement with reference~\cite{c9}.
The condition for a geometric resonance $\omega_B=2 \omega_Q$ thus yields trap aspect
ratios $\lambda_{1,2} =(\sqrt{125} \pm \sqrt{29})/\sqrt{72}$, i.e.~$\lambda_1\approx 0.683$ and $\lambda_2\approx 1.952$.

\Fref{fig:comparison} gives a comparison of the relative frequency shifts in the
TF limit calculated in reference~\cite{c9} using a hydrodynamic approach, and our
analytical results obtained using the Poincar\'e-Lindstedt perturbation theory.
Despite the good agreement, we observe small differences, which are due to the
fact that reference~\cite{c9} uses a parabolic variational ansatz for the condensate wave function, while
we use the Gaussian ansatz in equation~(\ref{eq:G}). We have confirmed that, when applied to the parabolic variational ansatz,
our perturbative approach yields frequency shifts in perfect agreement with reference~\cite{c9}.

\section{Resonant Mode Coupling}
\label{sec:MC}

In this section we study the phenomenon of non\-linearity-induced resonant mode coupling. As already
pointed out, even if a BEC system is excited precisely along the quadrupole or,
equivalently, the breathing mode, the emerging dynamics will lead to small
oscillations which initially involve only the corresponding mode, but then the other
collective mode will eventually step in, as well as higher harmonics of the two
modes and their linear combinations will appear. If the trap confinement  of the system allows a
geometric resonance, this could greatly enhance the mode coupling and speed up the emergence of those modes which are initially not excited, and therefore we
designate it as a resonant mode coupling. We focus on the experimentally most studied case of a repulsive two-body interaction, although all derived analytical results are valid also for the case of an attractive interaction. As effects of three-body interactions are usually small, and their main contribution is related to a stabilisation/destabilisation of the condensate, we focus on the emergence of resonant mode coupling due to geometric resonances.

To demonstrate this phenomenon, we use the perturbative solution of
equations~(\ref{eq:ur}) and (\ref{eq:uz}) with the initial conditions defined by
equations~(\ref{eq:quadrupoleinitial}), for which
the initial state coincides with the equilibrium with a small
perturbation proportional to the quadrupole mode eigenvector. The second-order
perturbative solution can then be written as
\begin{equation}
{\bf u}_0+\left(\begin{array}{c} A_{\rho Q} \\ A_{zQ}
\end{array}\right)  \cos \omega_Q t + \left(\begin{array}{c}
A_{\rho B} \\ A_{zB} \end{array}\right) \cos \omega_B t +\ldots \, ,
\end{equation}
where dots represent higher harmonics and the respective amplitudes are given by
\begin{eqnarray}
A_{\rho Q}&=&\varepsilon u_{\rho Q}+ \varepsilon^2 {\cal A}_{\rho
Q2}\frac{u_{\rho Q}^2}{\omega_Q^2}\, ,\\
A_{zQ}&=&c_1 A_{\rho Q}\, ,\\
A_{\rho B}&=&\varepsilon^2 {\cal A}_{\rho B2}\frac{u_{\rho Q}^2
(\omega_B^2-2\omega_Q^2)}{\omega_B^2 (\omega_B^2-4\omega_Q^2)}\,
,\label{eq:ArB}\\
A_{zB}&=& c_2 A_{\rho B}\, .
\label{eq:AzB}
\end{eqnarray}
Note that the absence of terms linear in $\varepsilon$ in expressions for
$A_{\rho B}$ and $A_{z B}$ is due to the initial condition, i.e. the fact that, initially, we only excite the quadrupole mode. The constants $c_{1,2}$ in the above expressions are
defined by equation~(\ref{eq:c12}), while ${\cal A}_{\rho Q2}$ and ${\cal A}_{\rho
B2}$ are calculated to be
\begin{eqnarray}
\hspace*{-5mm}
{\cal A}_{\rho Q2}&=&\frac{c_2 \gamma_\rho+ c_1 c_2 \alpha + c_1^2 c_2\beta
-\alpha - 4 c_1 \beta- c_1^2 \gamma_z}{3(c_1-c_2)} \ , \\
\hspace*{-5mm}
{\cal A}_{\rho B2}&=&\frac{-c_1^3 \beta + \alpha -  c_1 \gamma_\rho + 4 c_1
\beta - c_1^2 \alpha+c_1^2 \gamma_z}{c_1-c_2}\, ,
\end{eqnarray}
with $\alpha$, $\beta$, $\gamma_\rho$, $\gamma_z$ defined as
\begin{eqnarray}
\label{eq:alpha}
&&\alpha = \frac{3 p}{u_{\rho 0}^4
u_{z0}^2}+\frac{10 k}{u_{\rho 0}^6 u_{z0}^3}\, ,\quad
\beta=\frac{p}{u_{\rho 0}^3 u_{z0}^3 } +
\frac{3k}{u_{\rho 0}^5 u_{z0}^4 }\, ,\label{eq:alphabeta}\\
&&\gamma_\rho=\frac{6}{u_{\rho 0}^5}+\frac{6 p}{u_{\rho 0}^5 u_{z0}}
+\frac{15 k}{u_{\rho 0}^7 u_{z0}^2}\, ,\quad
\gamma_z=\frac{6}{u_{z0}^5} +
\frac{3p}{u_{\rho 0}^2 u_{z0}^4 } + \frac{6k}{u_{\rho 0}^4 u_{z0}^5 }\, .
\label{eq:grgz}
\end{eqnarray}

\begin{figure}[!t]
\centering
\includegraphics[height=5cm]{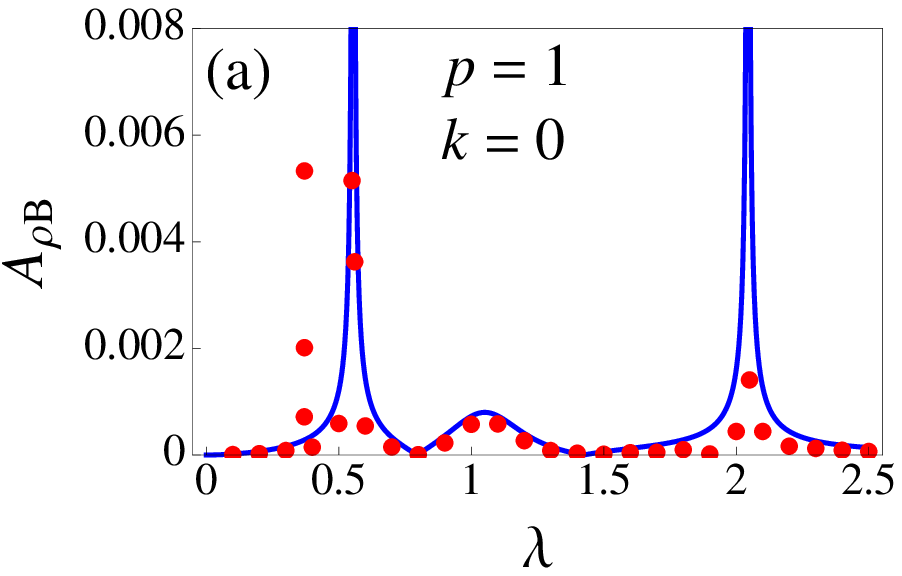}
\includegraphics[height=5cm]{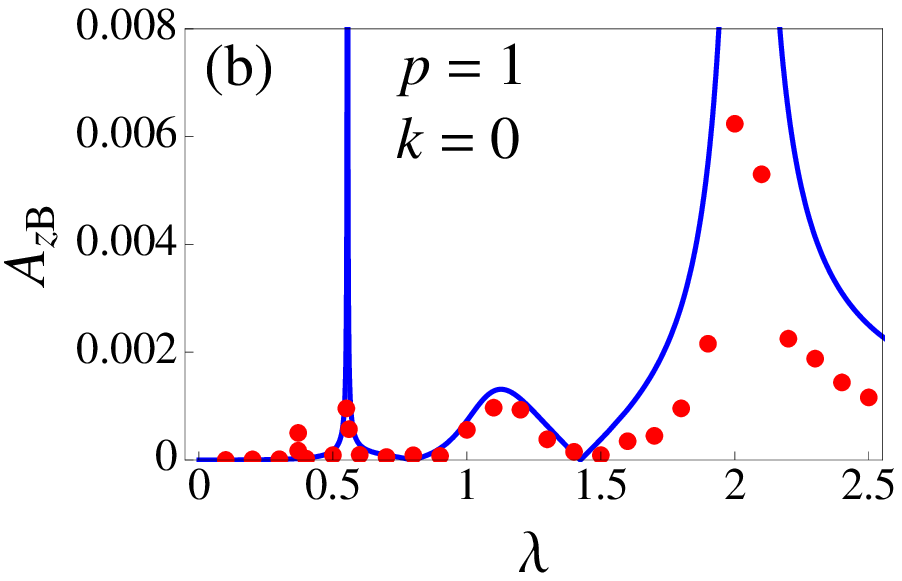}\\
\includegraphics[height=5cm]{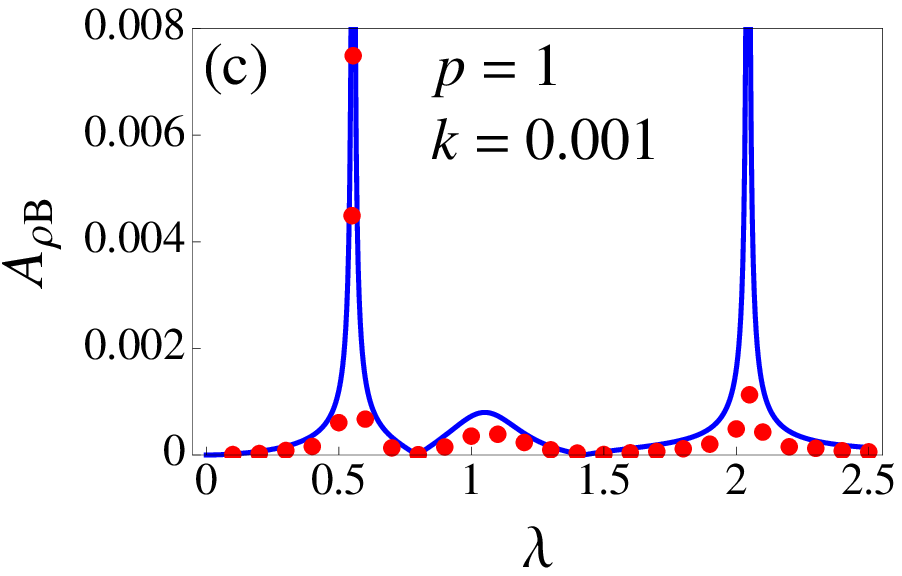}
\includegraphics[height=5cm]{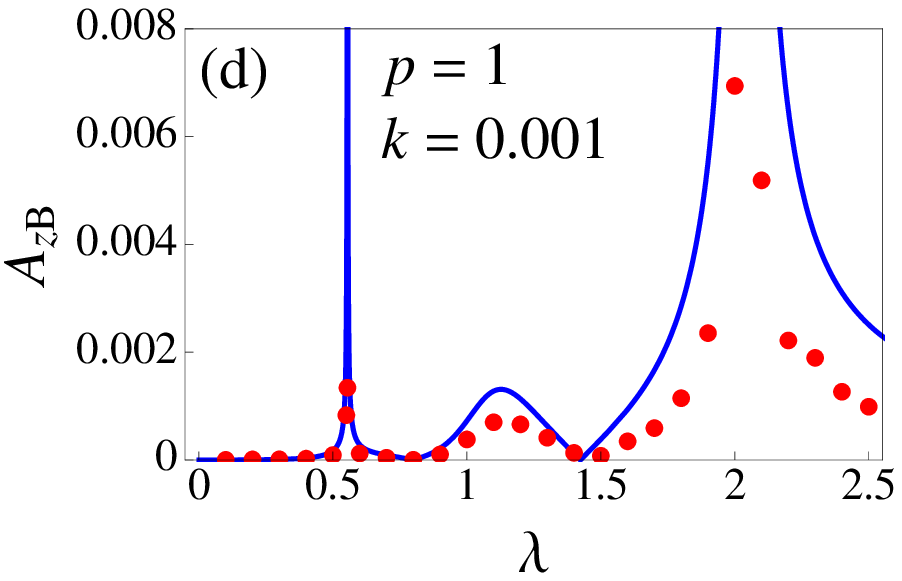}\\
\includegraphics[height=5cm]{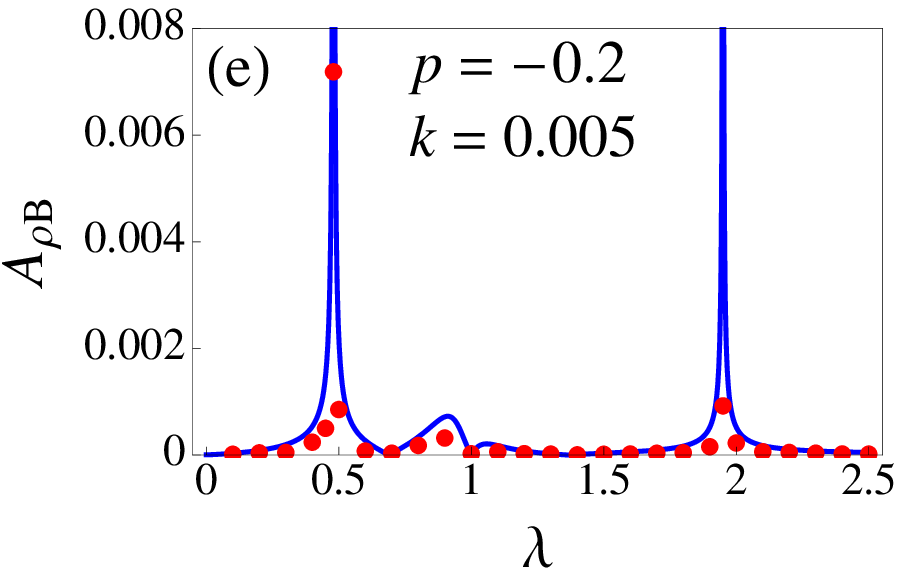}
\includegraphics[height=5cm]{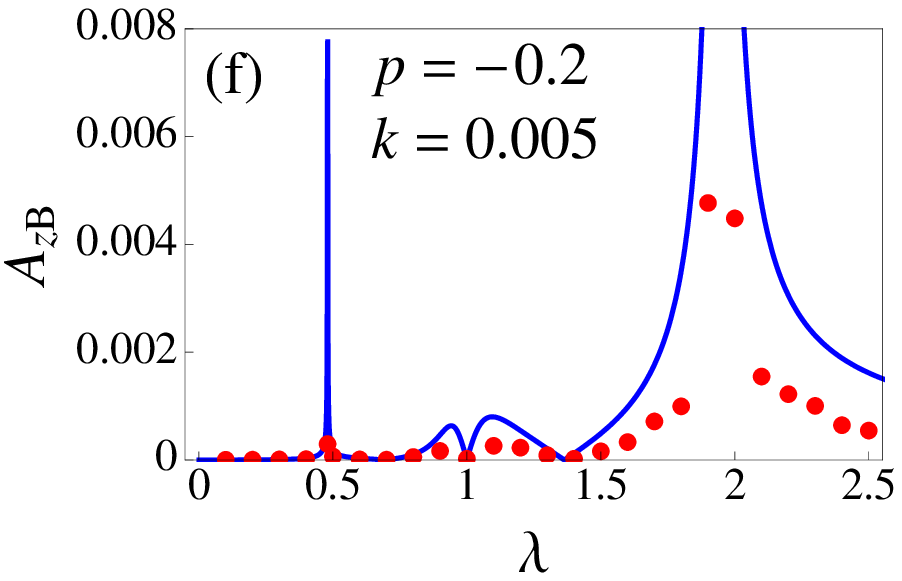}
\caption{Amplitudes of the breathing mode emerging in the second order of the
perturbation theory from BEC dynamics after initially only the
quadrupole mode is excited, given as functions of the trap aspect ratio
$\lambda$ for $\varepsilon=0.1$ and different values of two-body and three-body
interaction strengths $p$ and $k$. The amplitudes $A_{\rho B}$ and $A_{z B}$ from equations~(\ref{eq:ArB}) and ~(\ref{eq:AzB})
correspond to the radial and the longitudinal condensate width of the emerging breathing mode dynamics.}
\label{fig:AB}
\end{figure}

In \fref{fig:AB} we see the comparison of the derived analytical results, which emerge in the
second order, and corresponding numerical simulations for the amplitudes of the breathing mode. The numerical
results are obtained, as before, by extracting the amplitude of the breathing
mode from the Fourier excitation spectra of the system for each value of the
trap aspect ratio $\lambda$. The agreement is quite good, and we see again a resonant behaviour, which occurs at the same trap aspect ratios as for the frequency shift of the quadrupole
mode. From equations~(\ref{eq:ArB}) and (\ref{eq:AzB}) we actually see that the
resonances occur when the condition $\omega_B=2\omega_Q$ is satisfied, which is precisely
the same condition as for the frequency shift. This is not surprising, since amplitudes are expressed in terms of frequencies of the collective modes, and a resonant behaviour of the quadrupole mode frequency leads to resonances in the amplitudes for the same values of $\lambda$. Therefore, geometric resonances are not only reflected in the resonances of frequency shifts of collective modes, but also in the resonant mode coupling.

\begin{figure}[!t]
\centering
\includegraphics[height=5cm]{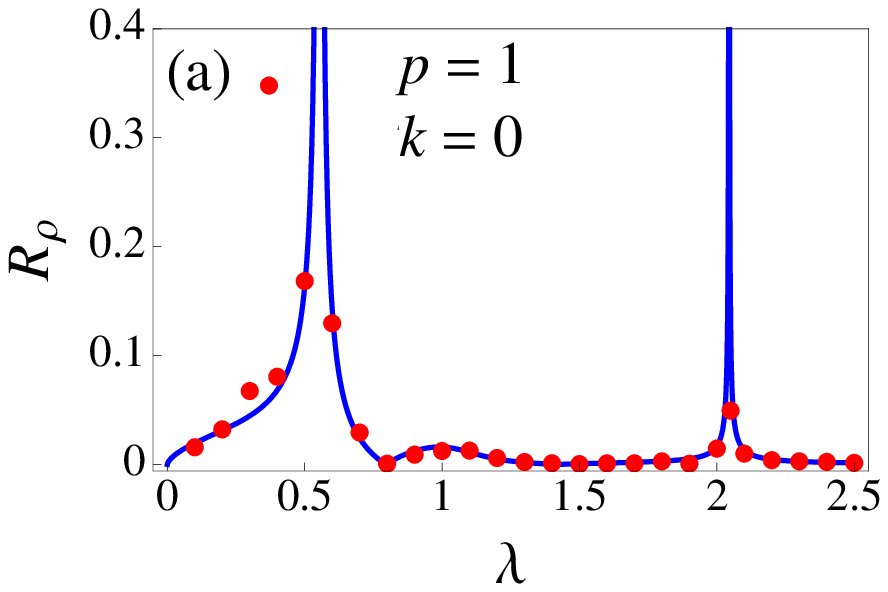}
\includegraphics[height=5cm]{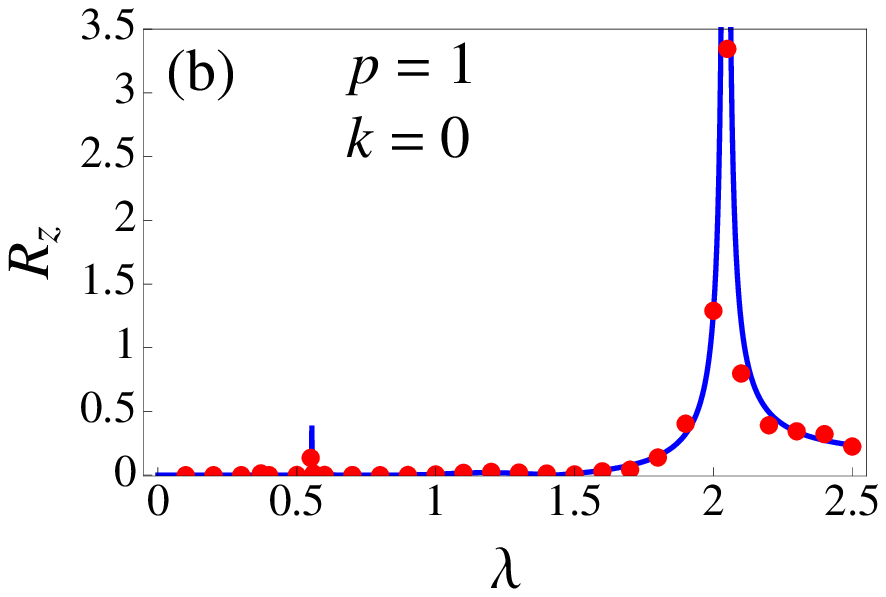}\\
\includegraphics[height=5cm]{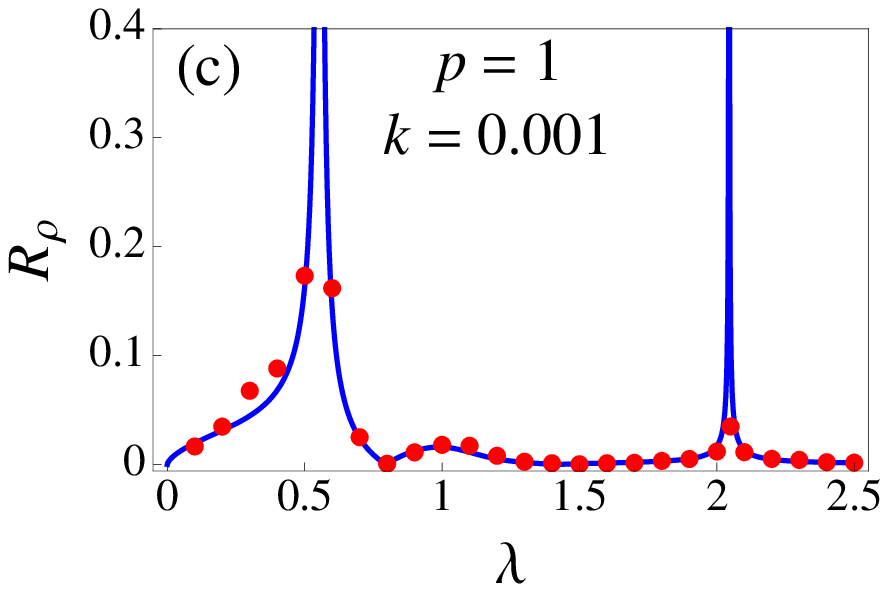}
\includegraphics[height=5cm]{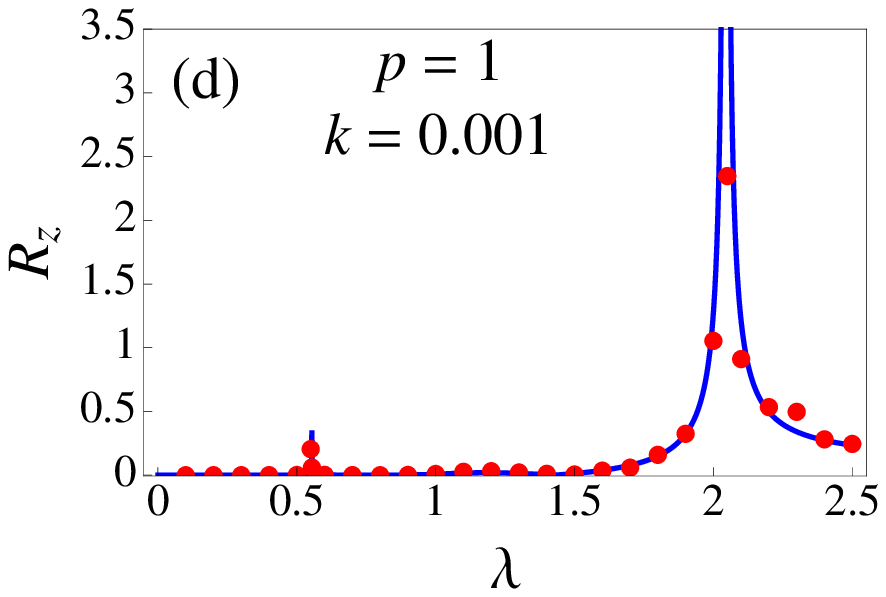}\\
\includegraphics[height=5cm]{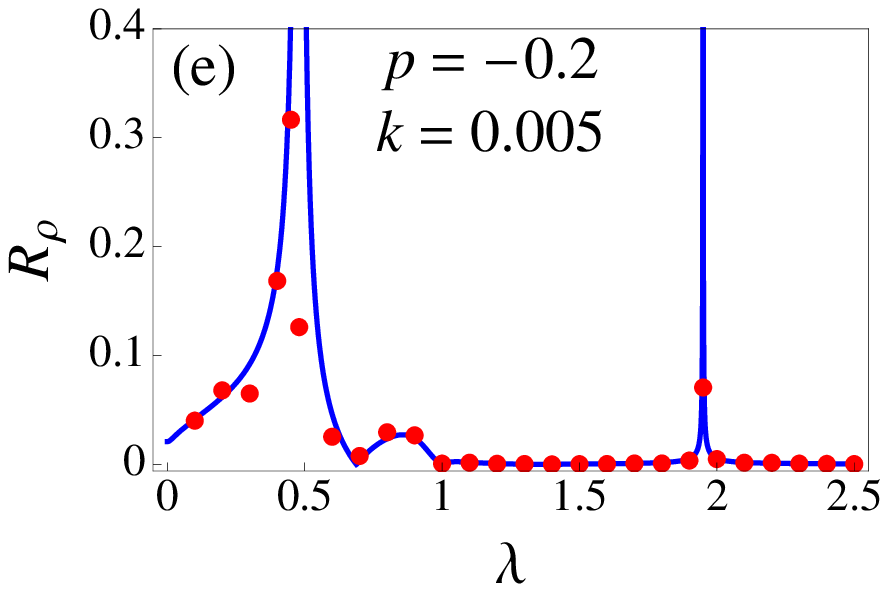}
\includegraphics[height=5cm]{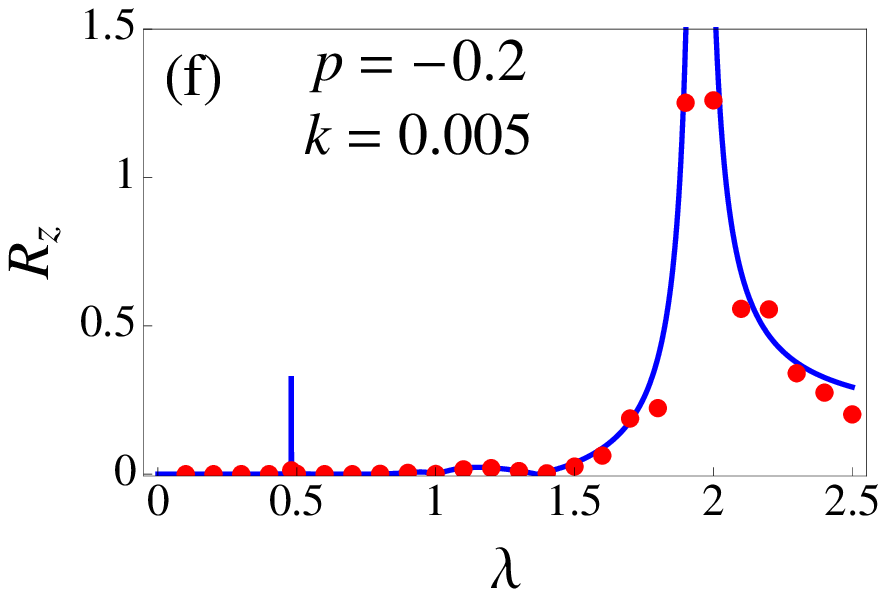}
\caption{Ratios of breathing and quadrupole mode amplitudes emerging in the
second order of the perturbation theory after initially only the quadrupole mode
is excited, given as functions of the trap aspect ratio $\lambda$ for
$\varepsilon=0.1$ and different values of two-body and three-body interaction
strengths $p$ and $k$. The quantities $R_{\rho}$ and $R_{z}$ from equations~(\ref{eq:RrB}) and (\ref{eq:RzB}) correspond
to ratios of amplitudes of the breathing and the quadrupole mode in the
radial and longitudinal condensate widths.}
\label{fig:RB}
\end{figure}

In addition to the absolute values of the breathing mode amplitudes, which are excited through
the resonant mode coupling, it is also interesting to look at their relative
values, compared to the quadrupole mode amplitudes, i.e.
\begin{eqnarray}
R_{\rho}&=&\frac{A_{\rho B}}{A_{\rho Q}}\propto
\frac{\omega_B^2-2\omega_Q^2}{\omega_B^2-4\omega_Q^2}\, ,\label{eq:RrB}\\
R_{z}&=&\frac{A_{z B}}{A_{z Q}}\propto
\frac{\omega_B^2-2\omega_Q^2}{\omega_B^2-4\omega_Q^2}\, .\label{eq:RzB}
\end{eqnarray}
\Fref{fig:RB} shows the comparison of analytical and numerical results for the
relative ratio of amplitudes of the resonance-excited breathing mode. Due to the geometric
resonances, we see that the trap aspect ratio can be tuned in such a way that
the resonant mode coupling excites the breathing mode with an amplitude far larger
than that of the quadrupole mode, which serves as the source of excitation.

Furthermore, from equations~(\ref{eq:ArB}) and (\ref{eq:AzB}) we see that, if
the geometry of the trap is tuned such that $\omega_B=\omega_Q\sqrt{2}$, then the
amplitudes of the breathing mode vanish simultaneously, i.e. $A_{\rho B}=A_{z
B}=0$. Although this is true only in the second-order perturbation theory, it still
represents a condition for a significant suppression of the resonant mode coupling. Therefore, the
tunability of the trap aspect ratio offers a unique tool for enhancing and suppressing the mode coupling in a BEC, which might be of broad
experimental interest.

In a similar way, we can initially excite only the breathing mode, which
corresponds to equations~(\ref{eq:ur})--(\ref{eq:uz}) with initial conditions defined
in
equations~(\ref{eq:breathinginitial}). The solution
in the second-order perturbation theory has again the form
\begin{equation}
{\bf u}_0 + \left(\begin{array}{c}
A_{\rho B} \\ A_{zB} \end{array}\right) \cos \omega_B t +\left(\begin{array}{c}
A_{\rho Q} \\ A_{zQ}
\end{array}\right)  \cos \omega_Q t  +\ldots \, ,
\end{equation}%
but now the respective amplitudes read
\begin{eqnarray}
A_{\rho B}&=&\varepsilon u_{\rho B}+ \varepsilon^2 {\cal A}_{\rho
B2}\frac{u_{\rho B}^2}{\omega_B^2}\, ,\\
A_{zB}&=&c_2 A_{\rho B}\, ,\\
A_{\rho Q}&=&\varepsilon^2 {\cal A}_{\rho Q2}\frac{u_{\rho B}^2
(2\omega_B^2-\omega_Q^2)}{\omega_Q^2 (4\omega_B^2-\omega_Q^2)}\,
,\label{eq:ArQ}\\
A_{zQ}&=& c_1 A_{\rho Q}\, ,
\label{eq:AzQ}
\end{eqnarray}
and the coefficients ${\cal A}_{\rho B2}$ and ${\cal A}_{\rho Q2}$ are given by
\begin{eqnarray}
\hspace*{-10mm}
&&{\cal A}_{\rho B2}=\frac{-c_1 \gamma_\rho - c_1 c_2 \alpha - c_1 c_2^2 \beta
+\alpha + 4 c_2 \beta + c_2^2 \gamma_z}{3(c_1-c_2)}\, ,\\
\hspace*{-10mm}
&&{\cal A}_{\rho Q2}=\frac{c_2^3 \beta -\alpha + c_2 \gamma_\rho - 4 c_2 \beta +
 c_2^2 \alpha - c_2^2 \gamma_z}{c_1-c_2} \, .
\end{eqnarray}
In this case, the ratios of amplitudes are given by
\begin{eqnarray}
R_{\rho}&=&\frac{A_{\rho Q}}{A_{\rho B}}\propto
\frac{2\omega_B^2-\omega_Q^2}{4\omega_B^2-\omega_Q^2}\, ,\label{eq:RrQ}\\
R_{z}&=&\frac{A_{z Q}}{A_{z B}}\propto
\frac{2\omega_B^2-\omega_Q^2}{4\omega_B^2-\omega_Q^2}\, .\label{eq:RzQ}
\end{eqnarray}

\begin{figure}[!t]
\centering
\includegraphics[height=4.82cm]{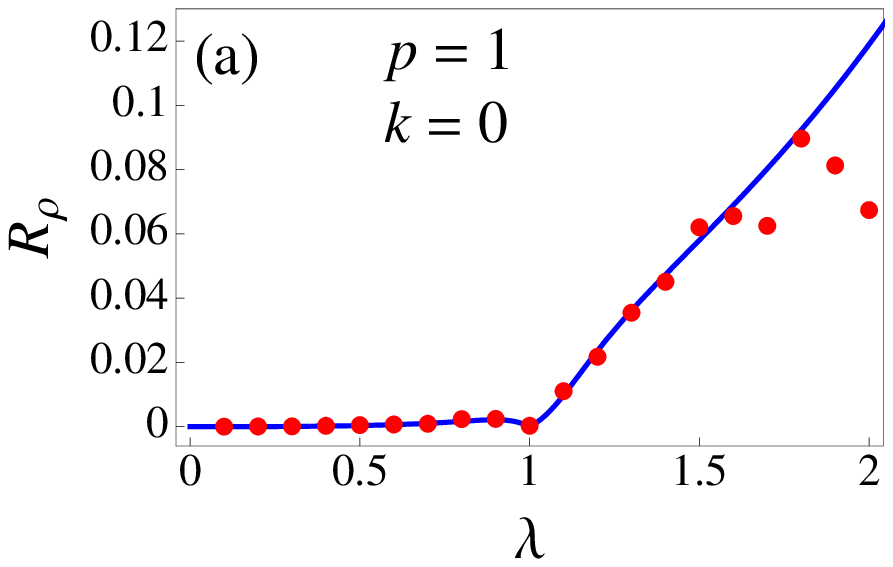}
\includegraphics[height=5cm]{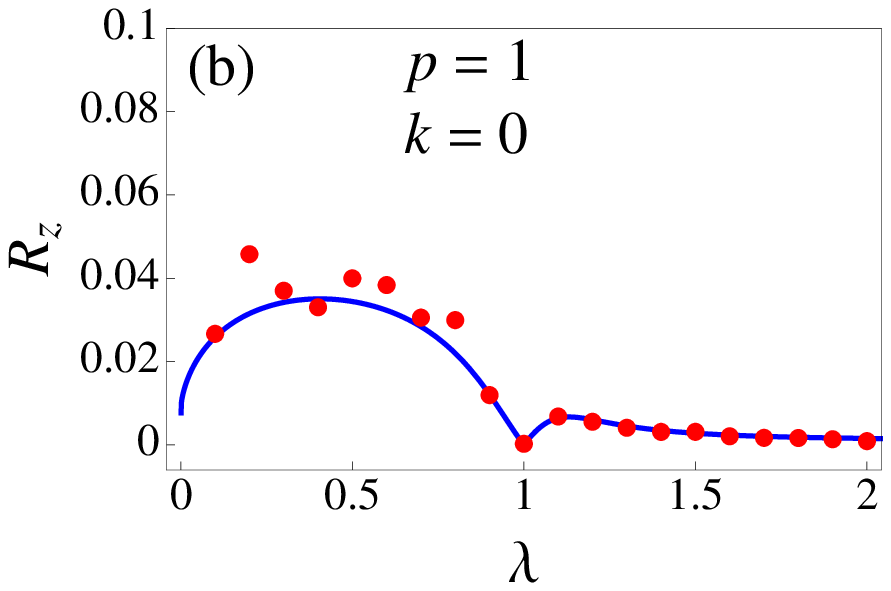}\\
\includegraphics[height=4.82cm]{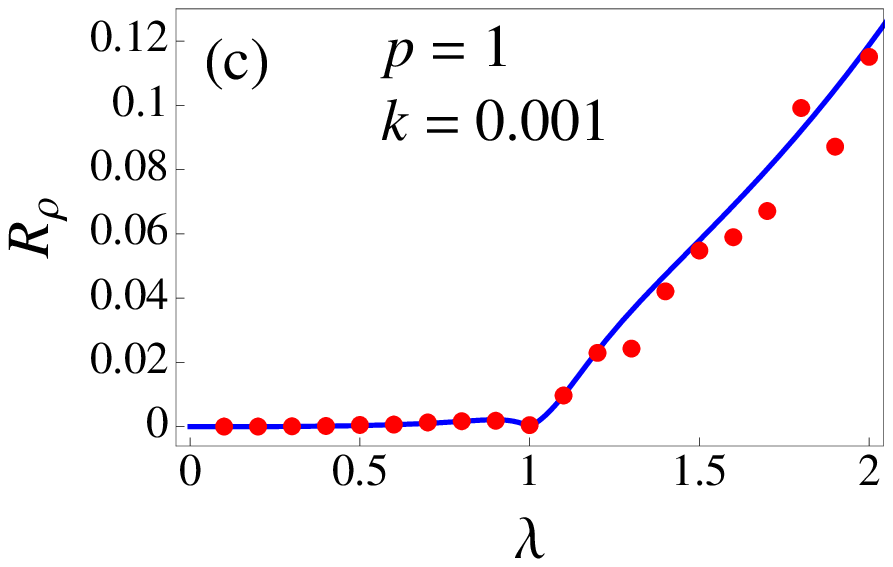}
\includegraphics[height=5cm]{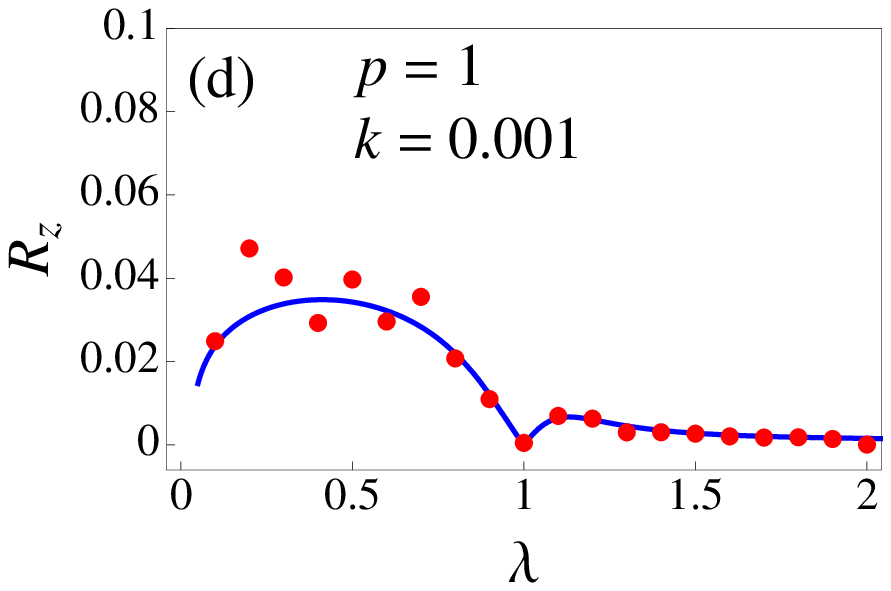}\\
\includegraphics[height=5cm]{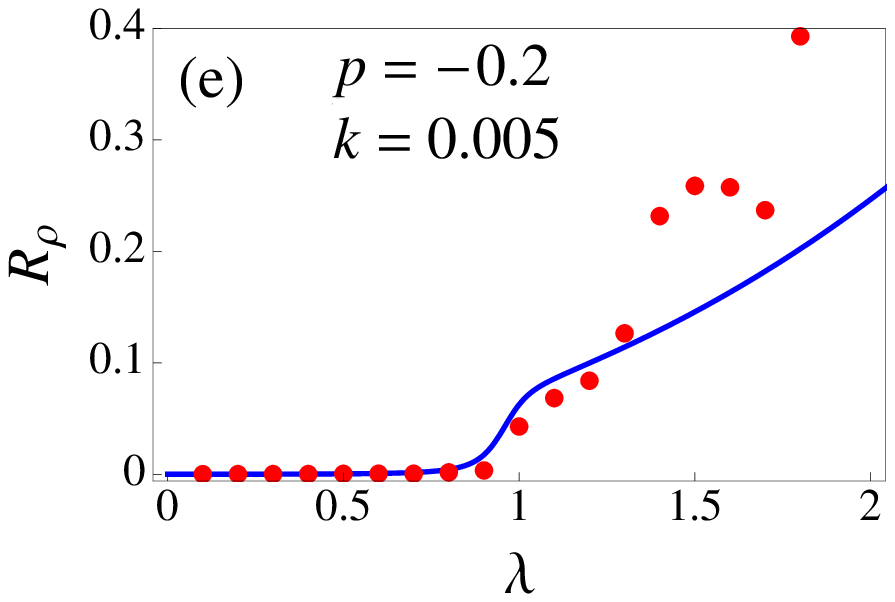}
\includegraphics[height=5cm]{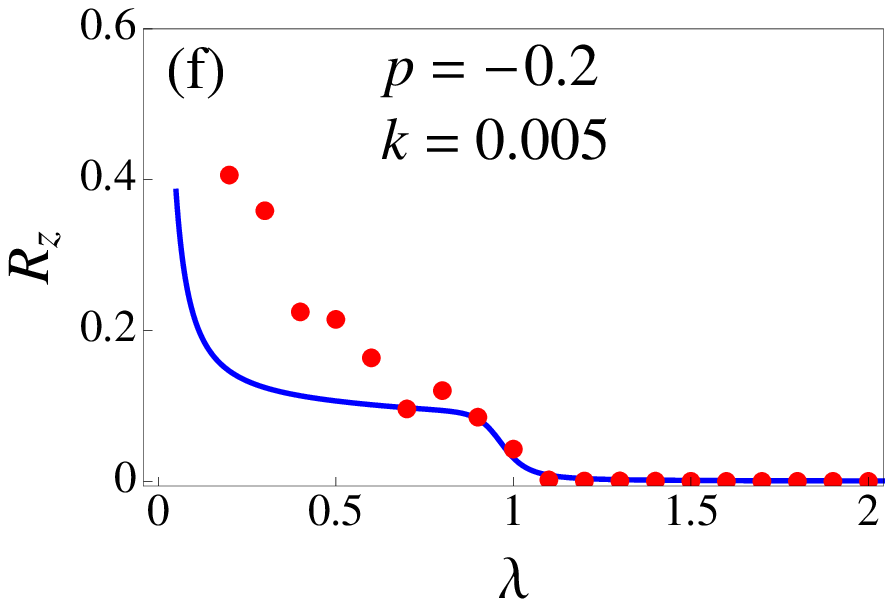}
\caption{Ratios of quadrupole and breathing mode amplitudes emerging in the
second order of the perturbation theory after initially only the breathing mode
is excited, given as functions of the trap aspect ratio $\lambda$ for
$\varepsilon=0.1$ and different values of two-body and three-body interaction
strengths $p$ and $k$. The quantities $R_{\rho}$ and $R_{z}$ from equations~(\ref{eq:RrQ}) and (\ref{eq:RzQ}) correspond
to ratios of amplitudes of the breathing and the quadrupole mode in the
radial and longitudinal condensate widths.}
\label{fig:RQ}
\end{figure}

Figure~\ref{fig:RQ} compares analytical and numerical results for the
mode coupling when initially only the breathing mode is excited, and then the
quadrupole mode emerges due to the mode coupling. As for the case of
the breathing mode frequency shift, there are no resonances, since
$\omega_B>\omega_Q$, and the resonance condition $2\omega_B=\omega_Q$ cannot be
satisfied, as is confirmed by the graphs. Therefore, the amplitudes do not experience resonances in this case, contrary to what we have observed in \fref{fig:RB}.
Again, this can be explained by the fact that amplitudes are functions of the breathing mode frequency, which does not experience any resonances, and hence
the same is valid for the corresponding amplitude.
Also, the condition $\omega_B\sqrt{2}=\omega_Q$ cannot be satisfied, and therefore the amplitude of
the quadrupole mode cannot be fully suppressed here, contrary to the case presented in \fref{fig:RB}. For a repulsive two-body
interaction in \fref{fig:RQ}(a)--(d) we see that the ratios $R_\rho$ and $R_z$
are below 10\%, and the mode coupling mechanism is not able to produce a
significant amplitude for the quadrupole mode. For the case of an attractive
two-body interaction in \fref{fig:RQ}(e)--(f), the ratio increases and the
generated quadrupole mode amplitude is stronger. Here the agreement between
analytical and numerical results is only qualitative, so that the perturbation theory
would have to be carried out to higher orders in the small parameter $\varepsilon$ in
order to improve the agreement.

\section{Conclusions}
\label{sec:Con}

We have studied the dynamics and collective excitations of a BEC for different
trap aspect ratios at zero temperature.
In particular, we have investigated prominent resonant effects that arise due
to two- and three-body interactions, and their delicate interplay.
We have discussed the stability of a condensate in an axially-symmetric
harmonic trap for the experimentally most relevant setups: repulsive and
attractive two-body interactions, attractive two-body and repulsive three-body
interactions, and attractive two- and three-body interactions.
We have shown that even a small repulsive three-body interaction is able to
extend the stability region of the condensate beyond the critical number of
atoms when the two-body interaction is attractive.

Using a perturbation theory and a Poincar\'e-Lindstedt analysis of a Gaussian
variational approach for the condensate wave function, we have studied in detail the relation between resonant
effects due to two- and three-body interactions, and the effects of the trap
geometry. Within the variational approach and the Poincar\'e-Lindstedt method,
we have analytically calculated frequency shifts and a mode coupling in order to identify geometric
resonances of collective oscillation modes of an axially-symmetric BEC.
We have also shown that the observed geometric resonances can be suppressed if
two- and three-body interactions are appropriately fine-tuned.

To verify the derived analytical results, we have used numerical
simulations, which provide detailed excitation spectra of the BEC dynamics.
We have numerically observed and analytically described several prominent
nonlinear features of BEC systems: significant shifts in the frequencies of
collective modes, generation of higher harmonics and linear combinations of
collective modes, as well as resonant and non-resonant mode coupling.
We have shown that, even if
the system is excited so that the perturbation corresponds initially to the
eigenvector of a particular mode, the nonlinear dynamics of the condensate will eventually
excite also other modes due to the mode coupling. The presence of geometric
resonances can significantly enhance this effect, as we have shown using the
developed perturbation theory.
All obtained analytical results are found to be in good agreement with the
numerical results. Furthermore, the results for frequency shifts are shown to
coincide with the earlier derived analytical results \cite{c9} within the
hydrodynamic approach in the Thomas-Fermi approximation.
In future work, we plan to extend the present analysis and also include the effects of quantum fluctuations \cite{Aristeu}.

\ack
This work was supported in part by the Ministry of Education, Science, and Tech\-nological Development of the
Republic of Serbia under projects No.~ON171017, NAD-BEC and NAI-DBEC, by DAAD - German
Academic and Exchange Service under projects NAD-BEC and NAI-DBEC, and by the European
Commission under EU FP7 projects PRACE-1IP, PRACE-2IP, PRACE-3IP, HP-SEE and EGI-InSPIRE.
Both A.~B. and A.~P. gratefully acknowledge a fellowship from the Hanse-Wissenschaftskolleg.

\end{document}